\documentclass[twocolumn]{aastex631}
\usepackage[varg]{txfonts}
\usepackage{graphicx}
\usepackage{graphics}
\usepackage{amssymb}
\usepackage{amsmath}
\usepackage[T1]{fontenc}
\usepackage[utf8]{inputenc}
\usepackage{color} 

\usepackage{dashrule}
\usepackage{nicematrix}
\usepackage{bigdelim}
\usepackage{booktabs}
\usepackage{threeparttable}
\newcommand{\source}{{ CGRABsJ0211+1051 }}

\newcommand{\swift}{{\it Swift}}
\newcommand{\fermi}{{\it Fermi}-LAT}
\newcommand{\xmm}{{\it XMM-Newton}}
\newcommand{\pd}[1]{\, \partial #1 \,}

\newcommand{\g}{\ensuremath{\gamma}}

\newcommand{\p}{^{\prime}}

\newcommand{\E}[1]{\times 10^{#1}}
\newcommand{\onehale}{\texttt{OneHaLe}}

\received{\today}

\submitjournal{ApJ}

\shorttitle{Curious Case of CGRaBS J0211+1051}
\shortauthors{Chandra S., et al.}

\begin{document}

\title{Curious Case of CGRaBS J0211+1051: Observational Evidence of Lepto-Hadronic Origin of High-Energy Emission?}

\correspondingauthor{Sunil Chandra}
\email{sunil.chandra355@gmail.com, schandra@prl.res.in}

\author[0000-0002-8776-1835]{Sunil Chandra}
\affiliation{Physical Research Laboratory, Navrangpura, Ahmedabad, 38009, India}
\affiliation{South African Astronomical Observatory, Observatory Road, Cape Town, South Africa}
\affiliation{Center for Space Research,
North-West University, Potchefstroom,
2520, South Africa}

\author[0000-0001-6890-2236]{Pankaj Kushwaha}
\affiliation{Department of Physical Sciences, Indian Institute of Science Education and Research, Mohali, 140306
Punjab, India}

\author[0000-0001-5430-4374]{Pranjupriya Goswami}
\affiliation{Université Paris Cité, CNRS, Astroparticule et Cosmologie, F-75013 Paris, France}
\affiliation{ZAH Landessternwarte, Universit\"at Heidelberg, Königstuhl 12, 69117 Heidelberg, Germany}

\author[0000-0001-5801-3945]{Michael Zacharias}
\affiliation{ZAH Landessternwarte, Universit\"at Heidelberg, Königstuhl 12, 69117 Heidelberg, Germany}
\affiliation{Center for Space Research, North-West University, Potchefstroom, 2520, South Africa}

\begin{abstract}
We present an extensive analysis of the multi-wavelength data of the low-synchrotron-peaked BL Lac object \source, which has been gathered over more than ten years with many observatories. Two major \g-ray flares have been observed during the \textit{Fermi} era: one in January 2011 and other in June 2019. During these events, \source\ was also bright in other energy bands. On the other hand, there are also examples of optical activity that do not exhibit any comparable $\gamma-$ray variability. Here, we study the temporal and spectral characteristics of the object in an attempt to understand the emission mechanisms operating in this source.
A peculiar feature in its spectrum is the X-ray domain, which is unusually soft considering its object class. Interestingly, the relatively soft UV and optical spectrum does not extrapolate well to the X-rays. To mimic the observed SEDs during quiescent and flaring periods, we use both a purely leptonic and a hadro-leptonic modeling approach to reproduce four broadband SEDs from various epochs. When taking into account the steep optical-UV spectrum, we find that the hadro-leptonic scenarios better explains the SEDs compared to the purely leptonic model. The hadro-leptonic interpretation of the two \g-ray flares suggests that \source\ could be both a potential neutrino emitter and TeV-bright (E$>$10~TeV). Thus, it may offer a unique test bed to check for hadro-leptonic contributions to the multi-messenger emission in blazar jets.

\end{abstract}


\section{Introduction}

Blazar is an extreme sub-class of Active Galactic Nuclei (AGNs), exhibiting a relativistic jet of plasma viewed within small angles to our line of sight (LOS; $\lesssim 10^{\circ}$). They are observationally characterized by a dominant non-thermal broadband variable emission (from radio to TeV bands) that exhibits a characteristic broad, dual-humped spectral energy distribution (SED). The low-energy emission component lies between radio to soft X-ray bands, exhibiting high and variable polarization at radio and optical ($\geq 3 \%$), and is attributed to the synchrotron processes in the relativistic jet \citep{1982ApJ...253...38U}. However, the high-energy emissions ranging from a fraction of GeV to tens of TeVs are explained by two very different but equally competitive models known as leptonic and hadronic scenarios. In the leptonic approach, the high-energy emission component is produced via inverse Compton up-scattering of low-energy seed photons by a distribution of highly energetic leptons (electrons/positrons). The origin of these seed photons can either be the synchrotron emission itself (Synchrotron self Compton, SSC; \citet{1989MNRAS.236..341G, Bloom1996, Sokolov_2004}) or photons from sources external to the jet, e.g., accretion disk, broad line region, torus, etc. (External Compton, EC). Several models have already been proposed demonstrating different external sources of seed photons taking part in high-energy generation in blazars \citep{1992A&A...256L..27D,1993ApJ...416..458D, 1994ApJS...90..923S, 2000ApJ...545..107B, 2009ApJ...704...38S, Agudo_2012}.  In the hadronic models, the relativistic protons traveling through the jet give rise to the gamma-ray emission by neutral pion decay ($p+\gamma\rightarrow\pi^0\longrightarrow\gamma$), proton synchrotron emission, and synchrotron emission from pair production ($p+\gamma\rightarrow\pi^\pm\longrightarrow\mu^\pm+\nu_\mu \longrightarrow e^\mp + \nu_e$)  either alone or in combination \citep{1999PASA...16..160M, 2001AIPC..558..700P,2000NuPhS..80C0810M, 2003APh....18..593M, 2005ApJ...630..186R,2020Galax...8...72C}.  \\

The first coincident detection of astrophysical neutrinos by IceCube from the direction of a flaring blazar TXS 0506+056 \citep{2018Sci...361.1378I} was a huge boost for the models supporting the hadronic origin of high-energy emission. The subsequent investigations of correlations between neutrino detection and  X-ray and Radio activities \citep{2024MNRAS.527L..26S, 2023MNRAS.526..942A, 2023ApJ...955L..32B, 2023ApJ...954...75A, 2023AAS...24111106D, 2015MNRAS.450.2658M} have strengthened the evidence of hadronic processes playing key roles in the high-energy emission in blazars.  In almost all kinds of hadronic models, extreme values of many a model parameters are required compared to the leptonic scenario [such as Lorentz boosting factor ($\Gamma$=$\delta$ $\sim$50),  high particle energy cutoffs ( $\gamma_{min}$ $\sim$ 10$^4$; $\gamma_{max}$$\sim$ 10$^9$) and magnetic field (B$\sim$10-50 Gauss)]. The proton synchrotron, for example, also demands jet power well exceeding the accretion power, indicating a super-Eddington accretion rate. It is found that Advection Dominated Accretion Flow (ADAF) disks can reach upto super-Eddington accretion rates.  However, expecting all the High-energy Peaking Blazars (HBLs) and Extreme HBLs to accrete in extreme conditions is a less favorable explanation.  Several hybrid models have been developed to tackle a bit of this tension and also to strengthen the roles of the ambient photons. The recent works by \citet{2023arXiv230713024R, 2023MNRAS.519.1396S} specifically focus on explaining the neutrino-associated multi-wavelength flaring activities of blazars in hybrid Lepto-hadronic scenarios. In most of the cases, lepto-hadronic, hadronic, and pure leptonic models equally explain the observed SEDs in blazars \citep{2013ApJ...768...54B}.

CGRaBS J0211+1051, first detected by the Energetic Gamma Ray Experiment Telescope (EGRET) instrument onboard the Compton Gamma Ray Observatory (CGRO) and later found to show featureless optical spectrum \citep{2008ApJS..175...97H},  is a BL Lac object associated with a host galaxy situated at the redshift of $z = 0.20\pm 0.05$ \citep{2010ApJ...712...14M}. The source was listed as an uncategorized blazar in the 1st \emph{Fermi} Catalog \citep{Abdo_2010}. It has now been classified as a low-synchrotron-peaked BL Lac or LSP from the study of optical polarization during its first flaring episode in the Fermi-LAT era in early 2011 \citep{2012ApJ...746...92C}. The polarization-based classification was firmly established in 2014 through its broadband SED \citep{2014ApJ...791...85C}. The modeling of SEDs observed at two different epochs (during the 2011 flare and quiescence in 2021), adopting a pure leptonic scenario, was performed by \citet{2022ApJ...931...59P}. The UV to X-ray spectra of \source, as typical of LBLs, exhibits a changeover between the synchrotron and inverse Comptonization processes, and hence enable a unique dataset to constrain the cutoff of the particle spectrum. Most of the LBLs are expected to be a TeV emitters, however, only a small subset have shown clear detection. In a study based on radio data from OVRO and Metsähovi Radio Observatories, \citet{2021A&A...650A..83H}, mainly an attempt to find a correlation between neutrino triggers with activities of CGRABS blazars in radio bands, connected a 390 TeV neutrino observed with IceCube during the radio variability of \source\, making this source an interesting object to explore more. The present study is a comprehensive investigation of the historic outbursts along with the prolonged $\gamma$-ray quiescence to uncover the emission processes and particle energization.  

This paper is organized as follows. Section~\ref{sec:data} discusses the observations and analysis of the multi-wavelength data. In Section~\ref{sec:results}, we discuss the SED model and important results. The last section describes our conclusions and the potential synergies with future observational facilities. 

\section{Observations and Data Analysis}\label{sec:data}

This work uses the publicly available (quasi)-simultaneous multi-wavelength data spanning over more than a decade between 2009 and 2021.

\subsection{ {\it Fermi}-LAT }
The photon and spacecraft data observed by \fermi\, corresponding to a search radius of 30$^\circ$ and energy range of 30 MeV to 500 GeV, are downloaded from the LAT data server\footnote{https://fermi.gsfc.nasa.gov/cgi-bin/ssc/LAT/LATDataQuery.cgi}. Fermitools package (v2.2.0 conda-release), and python package, {\tt fermipy}\footnote{http://fermipy-readthedocs.io/en/latest/}\citep[v0.6.6;][]{2017ICRC...35..824W} is extensively used to analyze the data. {\tt fermipy} facilitates various wrappers to perform the complete binned data analysis in an interactive manner. The initial set of parameters used for the events selection and primary cuts are: event class =128 (source), event type = 3 (front+back), region of interest (ROI) = 10$^\circ$, bin size = 0.1 pixels (for the map creation), maximum apparent zenith angle = 90$^\circ$. The instrument response functions (IRFs) P8R3\_SOURCE\_V2 is used throughout the analysis. The starting XML source model, created by {\tt fermipy} includes all the point-like and extended sources located within 25$^\circ$ radius of the location of \source\ as listed in the Fourth Fermi-LAT Source catalog \citep[4FGL][]{2020ApJS..247...33A}. It also includes the Galactic diffuse (gll\_iem\_v07.fits) and isotropic background emission (iso\_P8R3\_SOURCE\_V2\_v1.txt). For analyzing the data from different epochs, after ROI optimization and initial fitting, the point sources with negative test statistics (TS) are removed from the model. Following this, the spectral parameters of sources within 5$^\circ$ of \source\ are kept variable, while for others, they are fixed to the cataloged values with only free normalization parameter. The TS- and diffuse maps are generated for the verification of any possible GeV source (point and/or diffuse) not included in our model. Afrer a satisfactory optimization of the model, the best fit spectral parameters are estimated using the {\tt sed} procedure of {\tt fermipy} with an energy bin of 2 spectral points per decade. The 3-day or 5-day bins, chosen to have optimal number of significant flux estimations data during the flares without losing much of the timing information, are used to generate the lightcurves using the {\tt Lightcurve} procedure of the {\tt fermipy} package. In order to generate the lightcurves, the source model for \source\ which is ``LogParabola" as per the 4FGL catalog is approximated to ``PowerLaw" followed by the re-optimization of the model in order to reduced the number of free parameters for small time-bin calculations.      
     
The TS cutoff $\geq$ 25 (equivalent to $\geq$~5~$\sigma$ significance) and $\geq$ 16 ($\geq$~4~$\sigma$) are respectively used for generating final SEDs and lightcurves. For  the spectral and temporal bins not fulfilling above cuts are considered as lower-significance points, and hence 95\% upper limits are used. 
     
\subsection{Swift Observations}
The Neil Gehrels Swift Observatory, popularly known as \swift\footnote{\url{https://swift.gsfc.nasa.gov/}}, hosts two highly sensitive telescopes: a) X-ray Telescope \citep[XRT;][]{2005SSRv..120..165B}, and b) Ultraviolet/Optical Telescope \citep[UVOT;][]{2005SSRv..120...95R} enabling a total energy coverage of 2.3 eV to 10 keV. The HEASARC database\footnote{\url{https://heasarc.gsfc.nasa.gov/docs/archive.html}} enlists a total of 27 observations by \swift\ spanning over around 10 years of time between 2010 to 2020. All these observations are downloaded and analyzed using HEASoft package (v6.30\footnote{\url{https://heasarc.gsfc.nasa.gov/lheasoft/download.html}}) and adopting appropriate calibration updates\footnote{\url{https://heasarc.gsfc.nasa.gov/docs/heasarc/caldb}} for XRT (\citet{2005SSRv..120..165B}; calibration dated 15-09-2021) and UVOTA (\citet{2005SSRv..120...95R}; calibration dated 08-11-2021). Table \ref{tab:swiftdt} presents the metadata of all the X-ray data used in this work. 

\begin{table*}
\begin{center}
\caption{X-ray and UV observations from \swift\ and \xmm\ \label{tab:swiftdt}}
\begin{threeparttable}
\begin{tabular}{llllll}
\toprule
ObsID & Date-OBS & Time & Exposure & UVOT Filters & Flag \\
      & (UTC) & (MJD) & (s)   \\
\midrule
00039111001 & 2010-03-05T00:03:20 & 55260.0 & 844.1 & U, W2 & \hspace{0em}\rdelim\}{3}{*}[PPF2011] \\
00039111002 & 2010-06-13T15:32:45 & 55360.65 & 2002.8 & U \\
00041574001 & 2011-11-21T09:12:00 & 55886.38 & 941.5 & V, B, U, W1, M2, W2  \\
\hdashrule[0.5ex]{2.0cm}{1pt}{2pt} & \hdashrule[0.5ex]{3.0cm}{1pt}{2pt} & \hdashrule[0.5ex]{1.5cm}{1pt}{2pt} & \hdashrule[0.5ex]{1.0cm}{1pt}{2pt} & \hdashrule[0.5ex]{3.5cm}{1pt}{2pt} \\
00039111003 & 2011-01-25T20:16:25 & 55586.85 & 3920.8 & V, B, U, W1, M2, W2 & \hspace{0em}\rdelim\}{4}{*}[F2011]\\
00039111004 & 2011-01-28T23:10:22 & 55589.97 & 3863.3 & V, B, U, W1, M2, W2  \\
00039111005 & 2011-01-31T15:41:58 & 55592.65 & 3448.8 & V, B, U, W1, M2, W2 &   \\
00039111006 & 2011-02-03T17:34:16 & 55595.73 & 3933.2 & V, B, U, W1, M2, W2  \\
\hdashrule[0.5ex]{2.0cm}{1pt}{2pt} & \hdashrule[0.5ex]{3.0cm}{1pt}{2pt} & \hdashrule[0.5ex]{1.5cm}{1pt}{2pt} & \hdashrule[0.5ex]{1.0cm}{1pt}{2pt} & \hdashrule[0.5ex]{3.5cm}{1pt}{2pt} \\
03106474001$^\dagger$ & 2018-10-08T17:08:44 & 58399.72 & 257.2 & U, W1 \\
03106474002$^\dagger$ & 2018-10-09T18:37:43 & 58400.78 & 67.4 & U, W1 \\
00041574002 & 2018-11-23T20:44:13 & 58445.86 & 1713.1 & W2  & \hspace{0em}\rdelim\}{5}{*}[PreF2019]\\
00041574003 & 2018-12-14T06:04:39 & 58466.25 & 1176.2 & M2  \\
00041574004 & 2019-01-04T08:39:06 & 58487.36 & 1391.0 & W1   \\
00041574005 & 2019-01-25T08:18:55 & 58508.35 & 1967.9 & U, W2  \\
00041574006 & 2019-02-15T03:13:37 & 58529.14 & 2195.1 & W2   \\
03106067001$^\dagger$ & 2019-06-15T22:19:39 & 58649.93 & 189.8 & U, W1 \\
\hdashrule[0.5ex]{2.0cm}{1pt}{2pt} & \hdashrule[0.5ex]{3.0cm}{1pt}{2pt} & \hdashrule[0.5ex]{1.5cm}{1pt}{2pt} & \hdashrule[0.5ex]{1.0cm}{1pt}{2pt} & \hdashrule[0.5ex]{3.5cm}{1pt}{2pt} \\
00041574007 & 2019-07-01T03:14:20 & 58665.14 & 1460.9 & V, B, U, W1, M2, W2 & \hspace{0em}\rdelim\}{3}{*}[F2019]\\
00041574008 & 2019-07-02T02:52:32 & 58666.12 & 988.9 & V, B, U, W1, M2, W2  \\
00041574010 & 2019-07-04T04:18:32 & 58668.18 & 973.9 & V, B, U, W1, M2, W2  \\
\hdashrule[0.5ex]{2.0cm}{1pt}{2pt} & \hdashrule[0.5ex]{3.0cm}{1pt}{2pt} & \hdashrule[0.5ex]{1.5cm}{1pt}{2pt} & \hdashrule[0.5ex]{1.0cm}{1pt}{2pt} & \hdashrule[0.5ex]{3.5cm}{1pt}{2pt} \\
03106067002$^\dagger$ & 2019-10-08T12:43:19 & 58764.53 & 149.8 & U, W1 \\
03106474003$^\dagger$ & 2019-10-05T16:14:20 & 58761.68 & 149.8 & U, W1  \\
00083369001 & 2020-06-13T09:28:58 & 59013.4 & 829.1 & V, B, U, W1, M2, W2 & \hspace{0em}\rdelim\}{5}{*}[PostF2019]\\
00083369002 & 2020-06-14T17:17:01 & 59014.72 & 526.9 & V, B, U, W1, M2, W2  \\
03106067004 & 2020-06-20T08:41:12 & 59020.36 & 1113.8 & U, W1 \\
00083369003 & 2020-06-20T16:41:20 & 59020.7 & 449.5 & V, B, U, W1, M2, W2 \\
00083369004 & 2020-06-21T00:41:36 & 59021.03 & 551.9 & V, B, U, W1, M2, W2\\
03106067005$^\dagger$ & 2020-06-22T22:49:13 & 59022.95 & 754.2 & U, W1\\
03106067006$^\dagger$ & 2020-06-27T19:15:24 & 59027.8 & 64.9 & U, W1 \\
0861840101$^{\dagger\dagger}$ & 2021-02-04T22:58:40 & 59249.96 & 6.39$\times$10$^{4}$ & -- & \hspace{0em}\rdelim\}{1}{*}[Q2021]\\
\bottomrule
\end{tabular}
\begin{tablenotes}
\item[$^\dagger\equiv$] The exposure is too short, resulting in low SNR, and hence the X-ray data is unusable.
\item[$^{\dagger\dagger}$] XMM-Newton Observations\\
{\it Note: The group of \swift\ observations shown by different flags (e.g., PPF2011, F2011, PreF2019 etc.) are indicative of the observations grouped together to understand the X-ray spectral evolution before, during and after the major \g-ray activities. By definition they are different than F1, and F2 which are based on \g-ray flare duration and well incloses F2011 and F2019. Q2021 is equivalent to `Q' used for SEDs.  }   
\end{tablenotes}
\end{threeparttable}
\end{center}
\end{table*}

\subsubsection{XRT}  
The {\tt xrtpipeline} module, provided with the heasoft package, adapting the procedures prescribed by the instrument team \footnote{\url{https://www.swift.ac.uk/analysis/index.php}} is used to generate freshly calibrated cleaned events file. A default screening criteria is used for this stage. Due to limited exposure and poor signal-to-noise ratio (SNR), only PC mode observations are considered for this work. The default event grades 0-12 events are selected using Ftool {\it xselect} for generating source and background products. A circular area of radius 45 pixels around the target is used as the source region. Four source-free circular regions in the neighborhood, each of radius 45 pixels, are considered for the background. The required ancillary response matrix is generated using task {\it xrtmkarf} followed by {\it xrtcentroid}. The response matrix file provided with the CALDB distribution is used for further analysis. 

 \subsubsection{XMM-Newton} 

CGRABsJ0211+1051 was observed by XMM-Newton only once on 04/02/2021 (ObsID: 0861840101). We utilized the publicly available EPIC-PN and EPIC-MOS instrument data from the XMM-Newton Science Archive\footnote{\url{https://nxsa.esac.esa.int/nxsa-web/}}. To analyze this data, we used the XMM-Newton Science Analysis System (SAS) v21.0.0, the Updated Calibration Files (CCF), and followed the procedures outlined in SAS Data Analysis Threads\footnote{\url{https://www.cosmos.esa.int/web/xmm-newton/sas-threads}}. For EPIC-PN and EPIC-MOS data, we separately applied the SAS tasks {\tt emproc} and {\tt epproc} to generate calibrated event files, which include corrections for bad pixels and basic filtering. Additionally, we checked for periods of high background particle flaring activity\footnote{\url{https://www.cosmos.esa.int/web/xmm-newton/sas-thread-epic-filterbackground}}, thereby generating a good time interval (GTI) file that contains information on the good times that are free from soft proton flares using the {\tt tabgtigen} tool. We selected GTI with the background count rate threshold of ``RATE $<=$ 0.4" for PN and ``RATE $<=$ 0.35" for MOS. Subsequently, we generated cleaned event list files using the criteria ``FLAG == 0 \&\& PATTERN $<=$ 4" for PN and ``PATTERN $<=$ 12" for MOS. These filtered event lists were used in extracting the source image and source/background spectra. The source region was defined using a 30$\arcsec$ radius circle centered on the source, while the background was obtained from a source-free region employing two circular regions of 30$\arcsec$ on the same CCD. Furthermore, we utilized the SAS task {\tt epatplot} to detect potential pileup in the observation and to pinpoint suitable regions for source extraction. Using {\tt rmfgen} and {\tt arfgen} tasks, we generated response matrix files (RMF) and ancillary response files (ARF) for the extracted spectra. Subsequently, the spectra within the 0.3-10 keV energy range were grouped, ensuring a minimum of 25 counts for each spectral channel.

\subsubsection{UVOT}  
The observations from the UVOT aboard \swift\, performed in various optical and UV filters: V (5468 \AA), B (4392 \AA), U (3465 \AA), UVW1 (2600 \AA), UVM2(2246 \AA) and UVW2 (1928 \AA), listed in Table-\ref{tab:swiftdt}, are integrated with the {\it uvotimsum} task. The task {\it uvotsource} with a source region of $5\arcsec$ circle centred at \source is used to perform the analysis. A circular region of radius $45\arcsec$ covering a nearby source free patch of the image is used for  background estimates. The observed fluxes (in mJy) from all obsIds are then corrected for extinction according to the model described in \citet{1989ApJ...345..245C} before using them for the SEDs. 
    
\subsection{Optical Photometry}
Figure~\ref{fig:optlc} shows the lightcurve with optical monitoring data from the Catalina Real-Time Survey (CRTS)\footnote{\url{http://crts.caltech.edu/}}, which is publicly accessible via CSDR2\footnote{\url{http://nesssi.cacr.caltech.edu/DataRelease/CSDR2.html}}. 
The data from the Zwicky Transient Facility (ZTF)\footnote{\url{https://www.ztf.caltech.edu/}}, spanning over the duration of 2017-2021, are also used to extend the CRTS V band lightcurve beyond 2017, resulting in a decade-long time-series of \source.  The V band differential photometry from Steward observatory's archive \footnote{\url{http://james.as.arizona.edu/~psmith/Fermi/DATA}}  are also used to enrich the optical coverage. The observed differential magnitudes from the Steward Observatory are converted to the apparent magnitudes using the difference between two measurements performed within 4 hours of the same night.  Therefore, the corrected data may have systematics introduced because of the possible variability with timescales $\leq$ 4 hrs.

\subsection{Other archival ready to use data}
\begin{figure}
    \centering
    \includegraphics[width=0.5\textwidth]{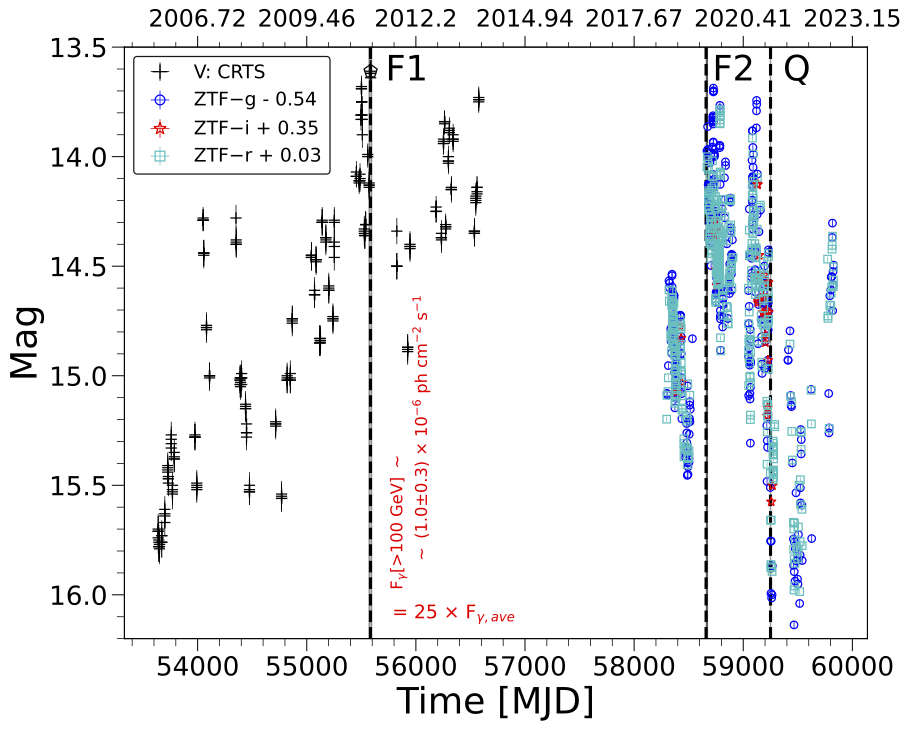}
    \caption{\label{fig:optlc}Long-term optical lightcurve showing strong variability throughout. CRTS and ZTF in the legend stand for photometry from Catalina Real-Time Sky Survey and Zwicky Transient Factory. The magnitudes in ZTF filters are shifted by a constant value to fit in the frame. The vertical dashed lines represent different epochs used for SED extraction.}
    \end{figure}

The published/cataloged data in different energy bands are also downloaded from SSDC SED Builder tool\footnote{\url{https://tools.ssdc.asi.it/SED/}}. This tool provides us with all publicly available, historical, ready-to-use data extracted from published resources or catalogs. The subset that we used in this work includes data from various catalogs of ground and space-based facilities, including OVRO, Plank, Wise, Fermi, etc. Epochs F1 (MJD 55570.0$-$55595.5), F2 (MJD 58653.4$-$58667.5) and Q (MJD 59248.0-59252.0), as indicated by the three vertical lines in Figure-\ref{fig:optlc}, represent the complete durations of the $\gamma$-ray flares (rise–peak–decay, as inferred from the \fermi\ light curve) in 2011 and 2019, and the period enveloping the \xmm\ observations and three ZTF observations during the lowest X-ray state recorded to date. This also corresponds to the lowest flux state observed in the UV/optical band.

\subsection{X-ray Spectral Fitting}

The X-ray spectra from the analysis of individual observation IDs revealed an extremely low count rate and, thus, very poor SNR. In order to increase the signal, the spectral products (source-, bkg-spectra, RMF, and ARFs) from these observations are combined using the {\tt addspec} tool to constrain the X-ray spectral behavior of the source. 

The energy filter ({\tt pha\_cutoff}) and grade filters on a few observations (the IDs as marked with asterisks in column 1 of the Table\ref{tab:swiftdt}) fail to provide any good events due to limited exposure times and hence are discarded. 

Combining only the neighboring observations with good events between 0.3-10 keV bands, a set of five merged spectra are extracted, representing the following time bins: pre \& post flare in 2010-2011 (PPF2011), flare in 2011 (F2011), quiescence in 2018 (PreF2019), flare in 2019 (F2019), and quiescence in 2020 (PostF2019). Refer to Table \ref{tab:swiftdt} for the details. 

The XRT observations in 2018 and 2020 reveal extremely faint states of the source, and we could only estimate the averaged flux over the 0.3-10 keV band.
 The remaining three spectra are modeled using absorbed powerlaw (M: {\tt TBaBs * Powerlaw}) with a fixed value of nH (=5.59 $\times$10$^{22}$ cm$^{-2}$, estimated using nH estimation tool in HEASoft tool). The {\tt WILM} abundance with {\tt VERN} photoelectric absorption cross-section was used for {\tt TBaBs} absorption model for the galactic extinction. Due to poor signal, {\tt cstat} was adopted as the statistics for the X-ray spectral modeling. 

\begin{figure*}
    \centering
    \includegraphics[width=0.36\textwidth, angle=-90]{SpecPlotSimFit_wilm_rev23.eps}
    \includegraphics[width=0.36\textwidth, angle=-90]{SpecPlotSimFit_wilm_tiedIndx_rev23.eps}\\
     \includegraphics[width=0.49\textwidth, trim={0 0 0 0},clip]{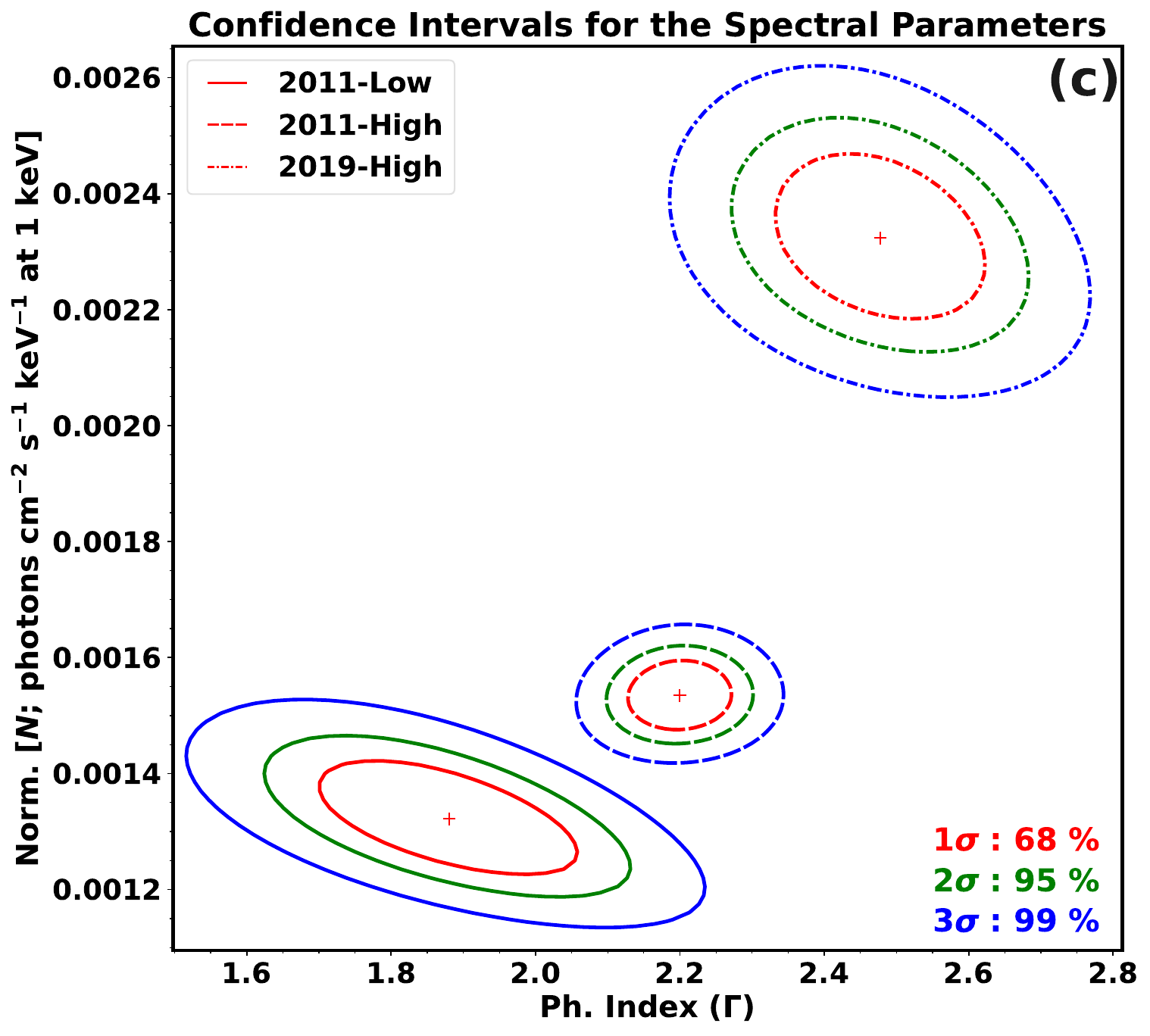}
    \includegraphics[width=0.49\textwidth, trim={0 0 0 0},clip]{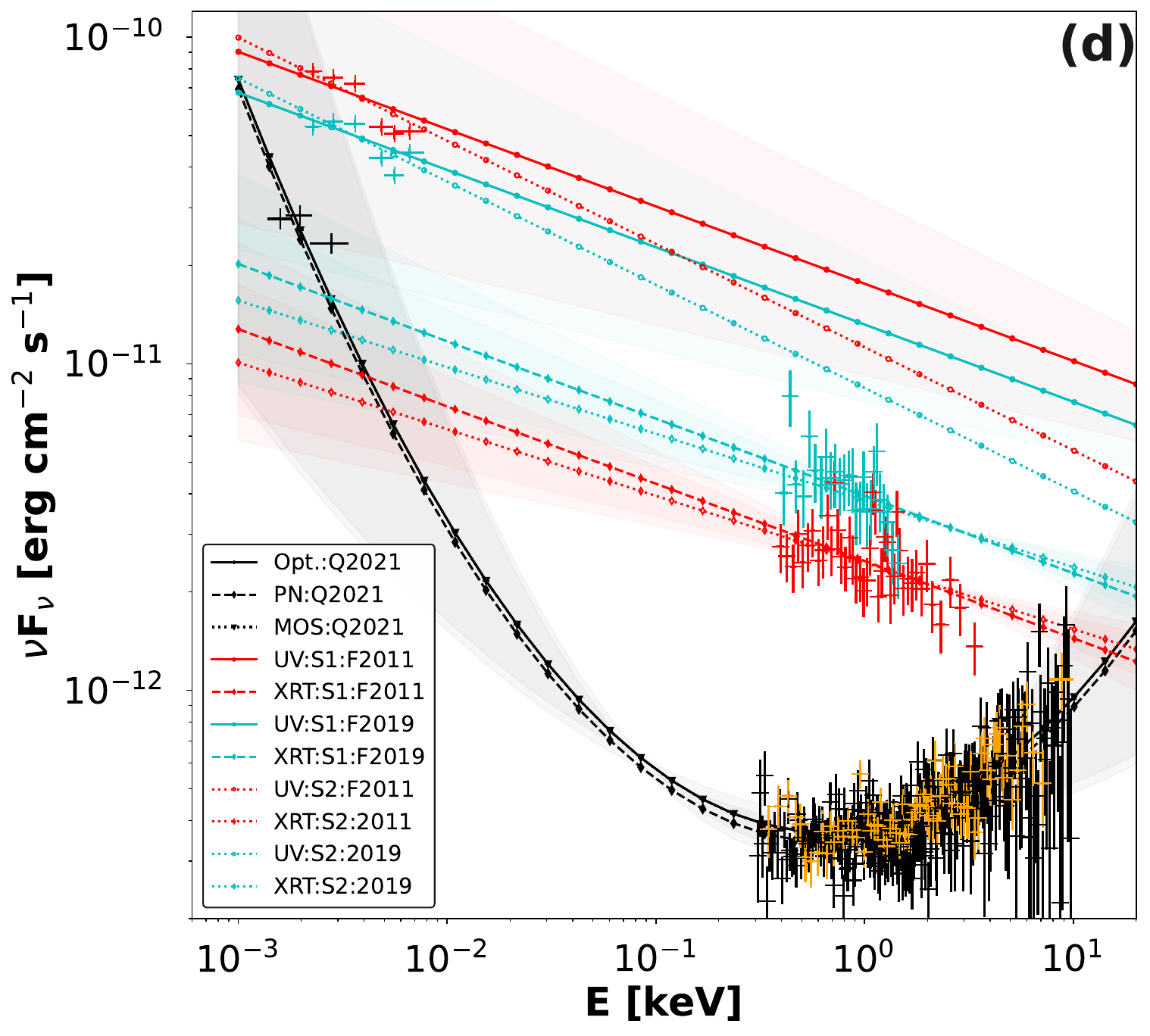}
    \caption{\label{fig:xspec_fits} {\bf~(a)}: X-ray Spectra fitted with simple absorbed powerlaw model ({\tt TBabs*powerlaw}). {\bf~(b):} Same as {\bf~(a)} but with $\Gamma$ tied to the one from 2011 observations. {\bf~(c):} The contour plots for $\Gamma-$N distribution showing confidence intervals for three different X-ray spectra. {\bf~(d):} Simultaneous X-ray and UV spectral modeling. The black data points refer to optical - X-ray data taken from December 09 - 15, 2021, representing a fainter quiescence state of the source. The red and cyan data are unfolded data points in UV-X-rays from F2011 and F2109. S1 in the legends refers to the test scenario-1 (ref. \S \ref{sec:uvxspecfit}): the spectral slope for UV and X-ray bands are the same, only norms are different, whereas S2 belongs to scenario-2: both norms and indices are kept free while modeling. The black X-ray data ('+') are PN spectra, whereas the orange X-ray data ('+') are from the MOS observations.
    }
\end{figure*}

 In order to extract reasonable spectral and flux constraints in 0.3-10 keV, the spectral modeling was performed simultaneously by 1) keeping the normalization and photon index as free parameters and 2) keeping photon indices tied together with the F2011. Using the first modeling approach, three confidence intervals were generated to quantify statistically significant spectral variations. The tool {\tt steppar} provided with {\tt XSPEC} was used to generate these contours. Figures~\ref{fig:xspec_fits}-(a),(b) respectively show the folded E$^2$F$_E$ spectra before and after tying the photon index. Fig.~\ref{fig:xspec_fits}-(c) shows the combined contour plots, representing the correlations between spectral changes and flux. The comparison of these contours clearly shows that the X-ray spectral index does not change significantly. The values obtained in F2019 are similar to the F2011 within 3$\sigma$ intervals, i.e., 2.2$\pm$0.01. 
 Given the observations are of limited exposures, instead of using individual spectral constraints, an average representative spectral shape, thus obtained, is a reasonable approximation. It is also noted that the three spectra show significant X-ray flux variations, with F2019 reaching a flux level of  $\sim$ 1.5 times F2011.

\subsection{Simultaneous UV/optical$-$X-ray Spectral modeling\label{sec:uvxspecfit}}
The averaged optical and UV photometry data, after the reddening corrections using prescription by \citep{1989ApJ...345..245C}, from the different time segments (referred to as F2011, F2019, and Q2021) are converted into spectral products (spectrum and response matrix) using {\tt ftflx2xsp} tool. These UV spectra are then simultaneously modeled with the respective X-ray spectra to understand the broadband spectral behavior. In order to model F2011 and F2019, we have started with the best-fit model and parameters from the previous subsection. Since optical-UV and X-ray both have softer spectra, we tested two 
scenarios: whether the extrapolated optical-UV spectrum explains X-rays with the same spectral shape (S1) or a different one (S2), as discussed below. The modeled spectra are shown in Fig.~\ref{fig:xspec_fits}-(d).

{\bf Scenario-1 (S1):} In the first scenario of modeling, the spectral shape ($\Gamma$) of UV and X-ray spectra are tied together, and the normalizations are independently free. In order to obtain reasonable statistics, UV and X-ray spectra of both F2011 and F2019 are fitted with identical spectral shapes. In this way, we can constrain the overall spectral shape between 0.001-5 keV and four normalizations representing UV and X-ray fluxes at two epochs. The solid red and cyan lines in Fig.~\ref{fig:xspec_fits}-(d) represent the modeled powerlaw spectra for the UV data during F2011 \& F2019, respectively. The dashed red and cyan lines represent the same but for X-ray spectra. These two modeled spectra, though, share the same spectral index but clearly do not reproduce each other when interpolated ahead. The UV normalizations during both epochs overproduce X-rays than the actually observed in identical spectral shape scenarios. 

{\bf Scenario-2 (S2):} In this setup, the optical-UV spectral shapes of F2011 and F2019 are tied together. The X-ray spectra during F2011 \& F2019 are also tied together separately. This way, through simultaneous fitting, we constrain two spectral shapes corresponding to UV ($\gamma_{uv}$) and other X-ray bands ($\gamma_{X}$), respectively. The dotted red and cyan lines passing through the UV and X-ray data points represent models for S2 (\ref{fig:xspec_fits}-(d)). In this scenario as well, the observed UV and X-ray spectra do not reproduce each other when extrapolated to the respective band. The fluxes in X-ray seem to be significantly lower than the model based on UV data. Despite poor data quality during F2011 and F2019, we deduce that under the adopted assumptions of powerlaw, UV, and X-ray emissions are driven by different mechanisms. 

Throughout the spectral study using \swift\ data, we have used cstat as a statistic. For the low state labeled as Q2021, the X-ray spectrum from {\it XMM}-Newton is modeled simultaneously with the averaged optical spectra from ZTF photometry performed during MJD 59248 to 59252. The joint spectra are best represented by a log parabola model showing an upturn towards hard X-rays ($\geq$2 keV). The thick black and dashed lines are model representations of the XMM and ZTF observations shown by black '+' symbols. 
It is clearly evident that Q2011 represents a deep quiescence in the 0.002-10 keV band. It indicates that we might have seen a similar upturn even during F2011 \& F2019 if it had similar total exposure as was adopted during Q2021. 

\section{Results and Discussion}\label{sec:results}
\subsection{Multi-Wavelength Lightcurves}

\begin{figure*}
    \centering
    \includegraphics[width=0.36\textwidth]{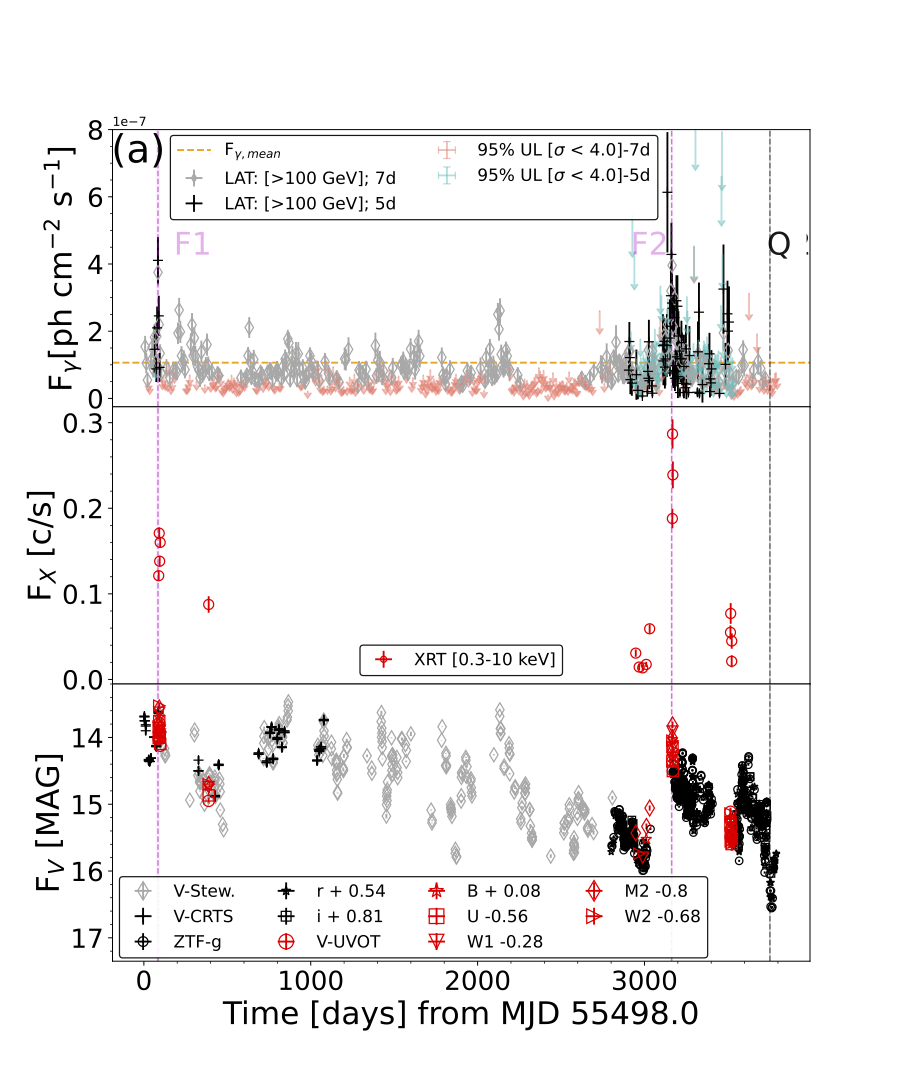}~
    \includegraphics[width=0.36\textwidth]{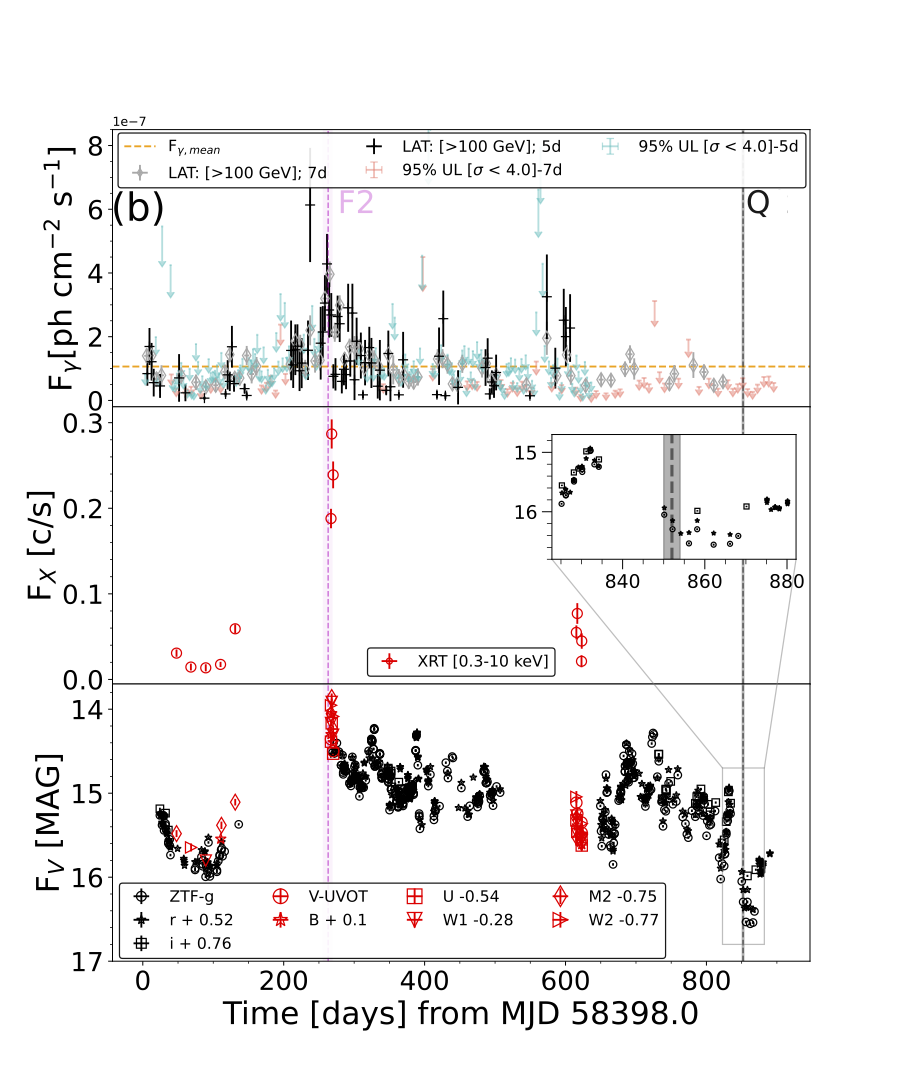}\\
    \includegraphics[width=0.36\textwidth]{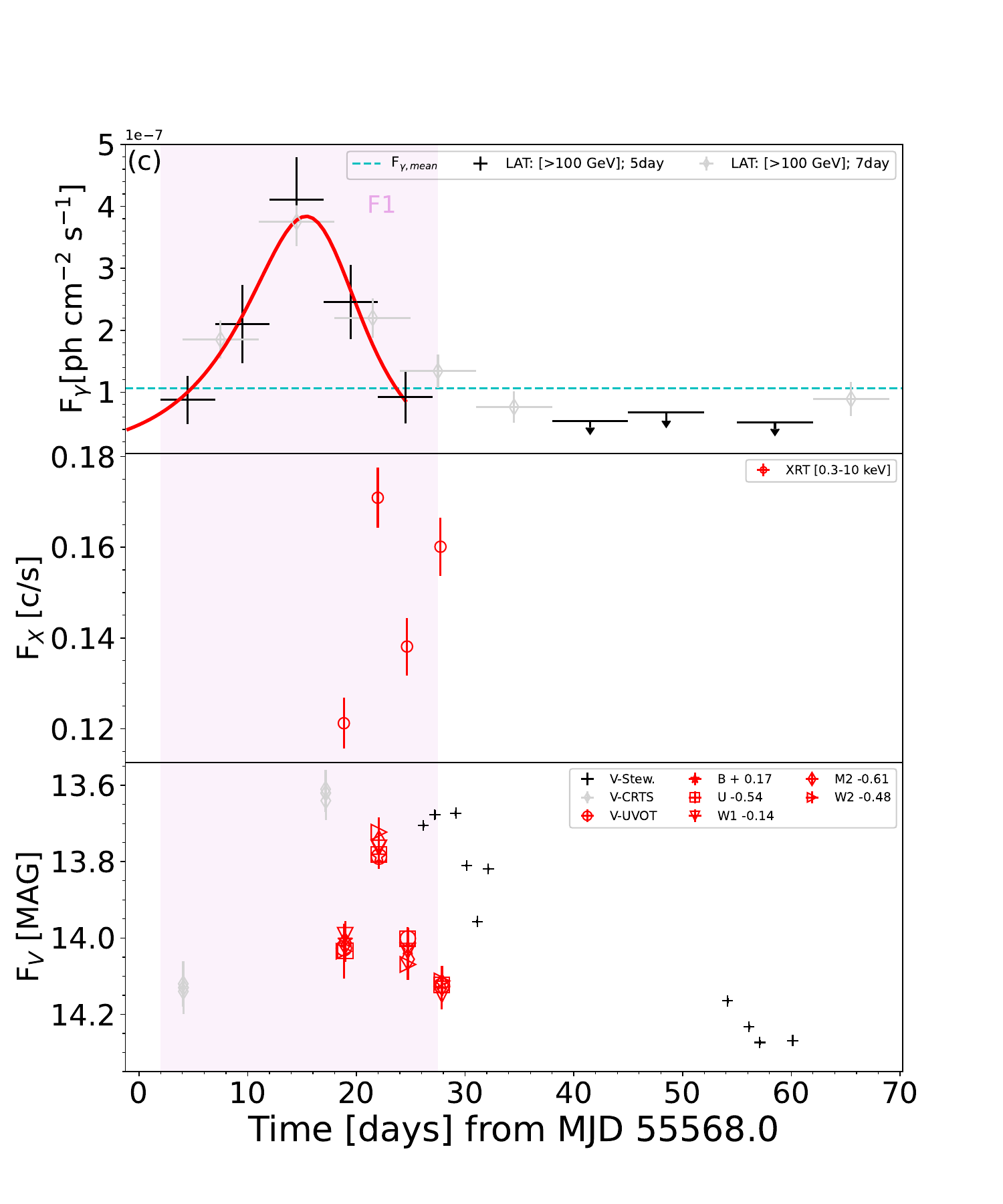}~
    \includegraphics[width=0.36\textwidth]{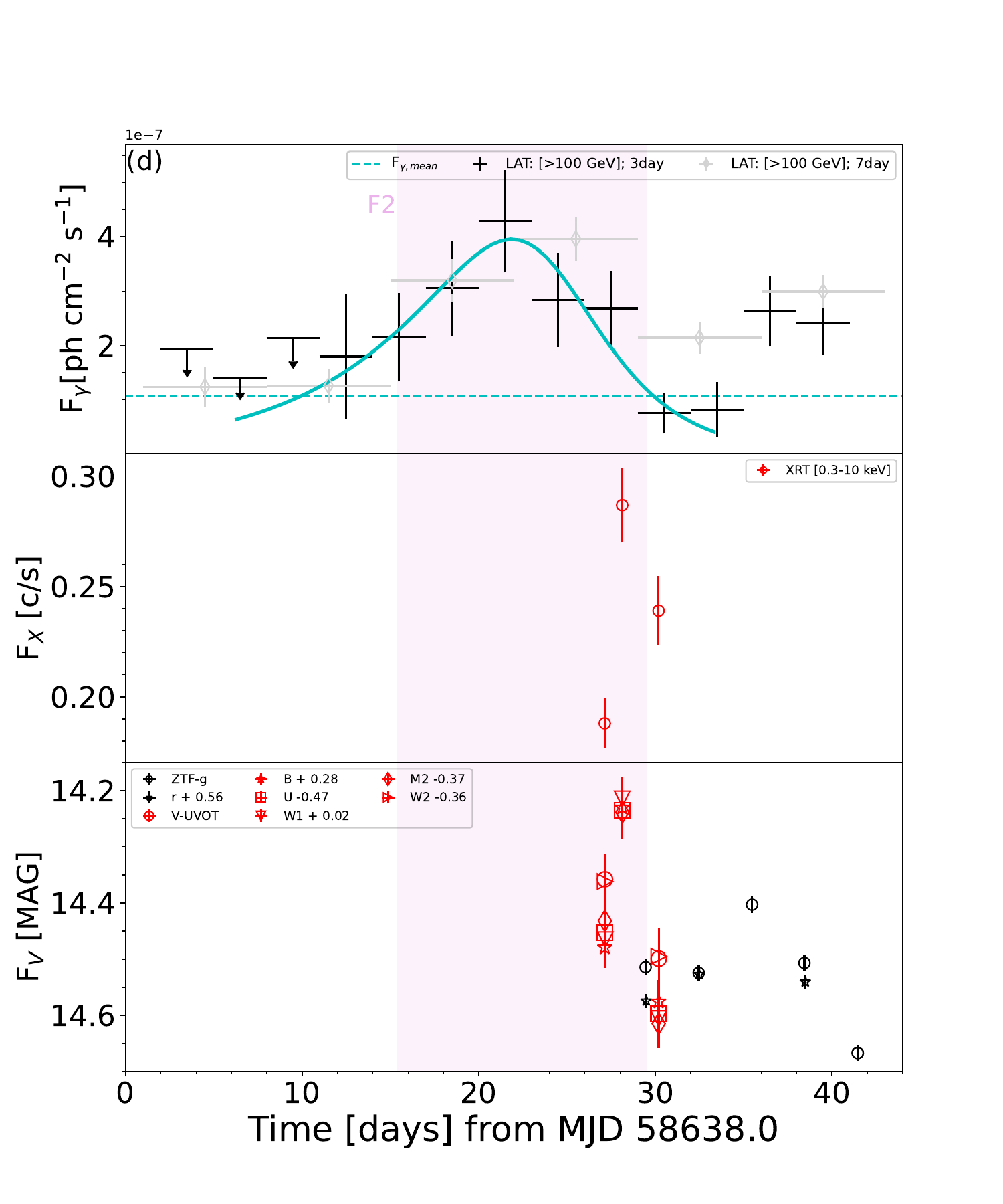}
    \caption{
    {\bf (a):} 10.5 year long multi-wavelength lightcurves of \source\ during UTC 2010-10-31 to 2021-03-27. Top panel: photon flux in $\gamma$ rays [$\geq$100 MeV] with black points being 5-day averaged fluxes derived from our analysis and grey points being 7-day averaged fluxes obtained from the LAT Lightcurve repository\footnote{\url{https://fermi.gsfc.nasa.gov/ssc/data/access/lat/LightCurveRepository/}}. Middle panel: count rate in X-rays [0.3-10.0 keV]. Bottom panel: magnitudes in optical and UV bands [{\it Swift}-UVOT, CRTS, Steward Observatory, and ZTF].  
    {\bf (b):} A zoom-in on 950 days enclosing F2 and Q. {\bf (c):} A zoom-in on F1. {\bf (d):} A zoom-in on F2. The red and blue solid lines, repectively represent the $\gamma-$ray flares F1 and F2 modeled using Equation-\ref{equ:flaremdl}. 
    }
    \label{fig:zmd_mwlc}
\end{figure*}

Figure~\ref{fig:zmd_mwlc}-(a) represents $\sim$10.5 year (MJD range: 55500 - 59300) long multi-wavelength lightcurves of \source comprising of gamma-ray from Fermi-LAT, X-ray (counts) from {\it Swift}-XRT, and optical-UV data from 
ground and space-based facilities (CRTS\footnote{\url{http://nesssi.cacr.caltech.edu/DataRelease/}}, zTF\footnote{\url{https://www.ztf.caltech.edu/ztf-public-releases.html}}, \swift-UVOT, and Steward\footnote{\url{https://james.as.arizona.edu/~psmith/Fermi/}}). 

The lightcurves show that  \source\ exhibited two major $\gamma$-ray outbursts. These are highlighted by vertical magenta lines, and are denoted F1 and F2, respectively, in Fig.-\ref{fig:zmd_mwlc}. Apart from these two outbursts, \source\ has shown minimal activity in the GeV band. The combined multi-band optical lightcurve suggests simultaneous activity around both $\gamma$-ray flares.  

Figs.~\ref{fig:zmd_mwlc}-(c),(d) show enlarged versions of the multi-wavelength lightcurve around F1 and F2. The absence of X-ray and optical data during the rise and peak of both $\gamma$-ray flares makes it difficult to estimate any time delay between the three bands. 

The $\gamma$-ray flares span over $\sim$15 days during F1 and $\sim$20 days during F2. Both the \g-ray flares were characterized by a function of the sum of two exponential defined as \citep{2010ApJ...722..520A}:
\begin{equation}
    F(t) = 2F_0\big(e^\frac{t_0 - t}{T_r} + e^\frac{t - t_0}{T_d}\big)^{-1}
    \label{equ:flaremdl}
\end{equation}
where t$_0$ and F$_0$ refers to the time of the peak flare and amplitude of the flare, respectively. T$_r$ and T$_r$ are rise time and decay time of the flare. The results of the fit are given in Table \ref{tbl:flrfitmdl} and shown in Figures \ref{fig:zmd_mwlc}. We also estimated a parameter $\xi$ = $\frac{T_d - T_r}{T_d + T_r}$, which describes the symmetry of the flares \citep{2010ApJ...722..520A}. For the F1, the flare seems to be symmetrical, however, as suggested by $\xi$ there is a slight asymmetry in F2.

\begin{table}
    \centering
    {\footnotesize \begin{tabular}{l|c|c|c|c|c}
    \hline
      & t$_0$ & F$_0^\dagger$ & T$_r$ & T$_d$ & $\xi$ \\
        &  (MJD)  &  & (d) & (d) & \\
        &         & & & &  \\
     \hline
        F1 & 55584.4$\pm$1.4 & 3.75$\pm$0.37 & 5.95$\pm$0.69 & 3.79$\pm$0.93 & -0.22\\
        F2 & 58661.5$\pm$2.2 & 3.71$\pm$0.63 & 7.02$\pm$1.54 & 3.38$\pm$1.07 & -0.35 \\\hline 
    \end{tabular}}
    \caption{Best fit parameters to characterize the gamma-ray flares using equation \ref{equ:flaremdl}.\\$^\dagger$: the peak flux in unit of 10$^{-7}$ ph cm$^{-2}$ s$^{-1}$.}
    \label{tbl:flrfitmdl}
\end{table}

During both the flares, the sparsely spaced time-series hint that X-ray and optical emission possibly show simultaneous shorter-time scale variations (F1: $\Delta$t $\sim$ 3 days \& F2: $\Delta$t $\sim$ 1.5 days) during the decreasing phase of the $\gamma$-ray flares. After 2019 peak, the well-sampled optical light curve shows many re-brightening phases superposed on an overall declining profile. The time scales are estimated as half of the total duration of rise-peak-decay of flux points in the respective energy bands. \\  

Overall the decade-long optical lightcurve shown in Fig.~\ref{fig:optlc} reveals that \source\ exhibits strong variations on various time scale. Several flare-like features can be seen, which have no or very feeble counterparts at GeV. The F1 is associated with the brightest ever state in the V band, measured within $\sim$ 3 days of the $\gamma-$ray flare peak. The general trend in the V-band (ref. figure \ref{fig:optlc}) shows a systematic brightening of the source since $\sim$ MJD 54600.0 (i.e., 14 May 2008) with  many superposed dimming and brightening events on different timescales (e.g. 0.8-0.9 mag over few months to as short as within 4-days in ZTF bands), reaching peak during F1 while F2 followed a systematic dimming phase. After F2, again dimming and brightening phases superposed on an overall declining envelop is apparent. 
 
It is also noticeable that as a qualitative measure the optical flux changes more frequently during a $\gamma-$ray flare than at other epochs. Also, during broad optical activities, the $\gamma-$ray flux remains low, close to the mean GeV flux estimated for over 10.5 years. This hints at a scenario where most of the large time-scale optical variations have at most weak $\gamma-$ray counterparts. In contrast, both $\gamma-$ray flares of \source\ have strong optical counterparts. The sparse X-ray lightcurve shows similar trend to that of optical, including a potential mirroring of the rapid optical variability during $\gamma-$ray flares in the X-ray domain. The optical and $\gamma-$ray lightcurves show that Q corresponds to a faint state in both bands with only upper limits (7-day averaged) in gamma-ray while the g-, i-, and r-band magnitudes are close to the faintest observed magnitudes in these bands over the duration of this study.  

As mentioned before, the \g-ray variability typically shows concurrent optical flares, while the opposite is not necessarily true. From these two observed scenarios of optical/UV and $\gamma-$ray emissions, we can qualitatively emphasize that different locations of the emission region may play a role. We can safely say that most of the time, when we see only optical/UV flares, the variability arises either from 1) close to the central engine where the $\gamma-$rays are attenuated by photon-photon pair production or 2) from very far from BLR where the Comptonization is too weak. Considering the participation of the relativistic protons in the processes behind high-energy emission brings a number of possibilities to explain the observed behavior. If the hadronic emission is dominated by the p-$\gamma$ interactions, then the role of ambient external photon densities in IR/optical/UV/soft-X rays becomes very important. If the emission indeed originates in different locations, one can expect different variability patterns in the optical and \g-ray lightcurves. A rigorous timing study (cross-correlation and lags) would be helpful in order to extract more details about the long-term variability behavior of \source. However, the lack of well sampling prohibits further investigation in this direction.  
      
\subsection{Spectral Energy Distribution (SED)\label{sec:sedmod}}

\begin{figure*}
    \centering
    \includegraphics[width=0.40\textwidth]{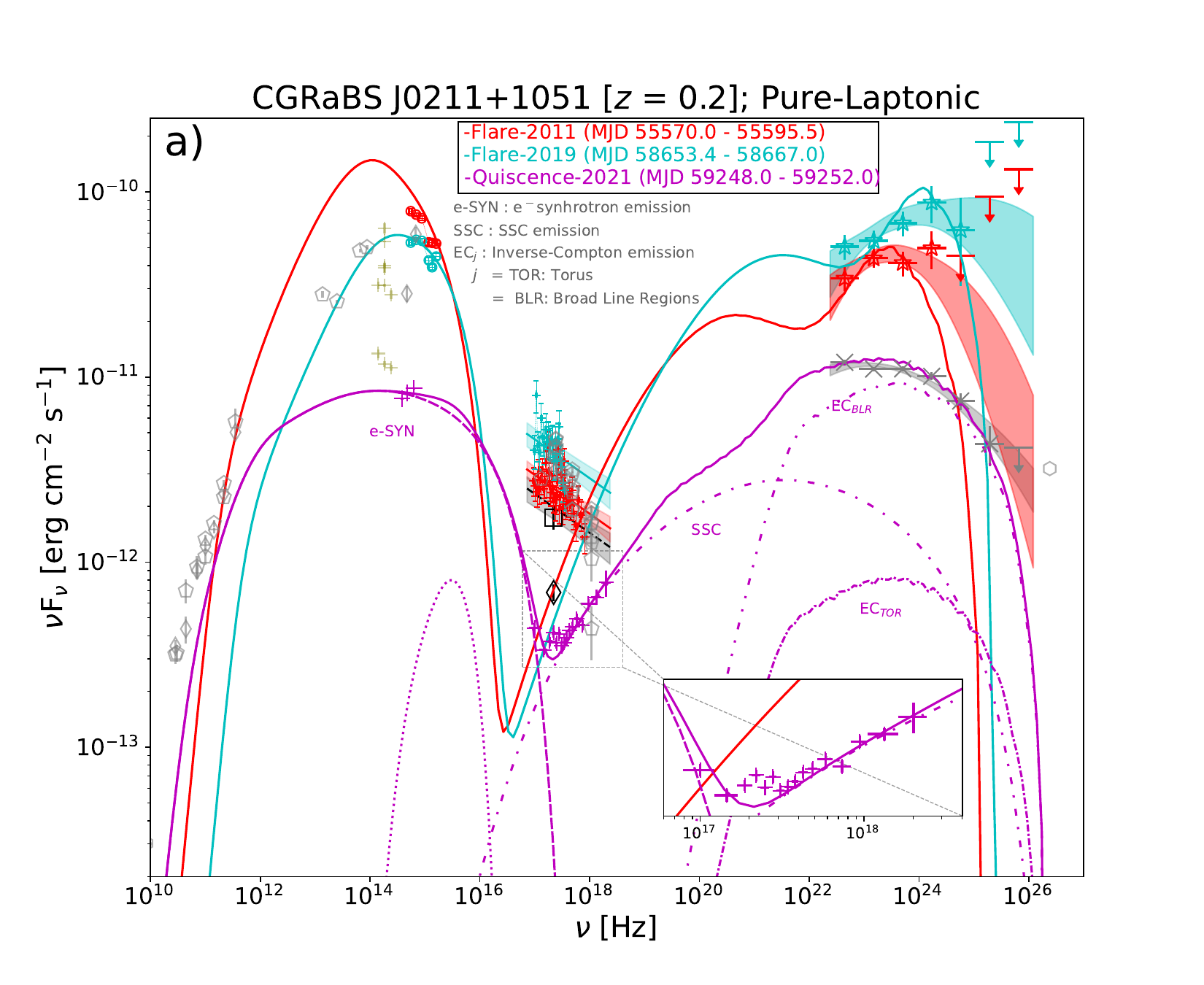}
    \includegraphics[width=0.40\textwidth]{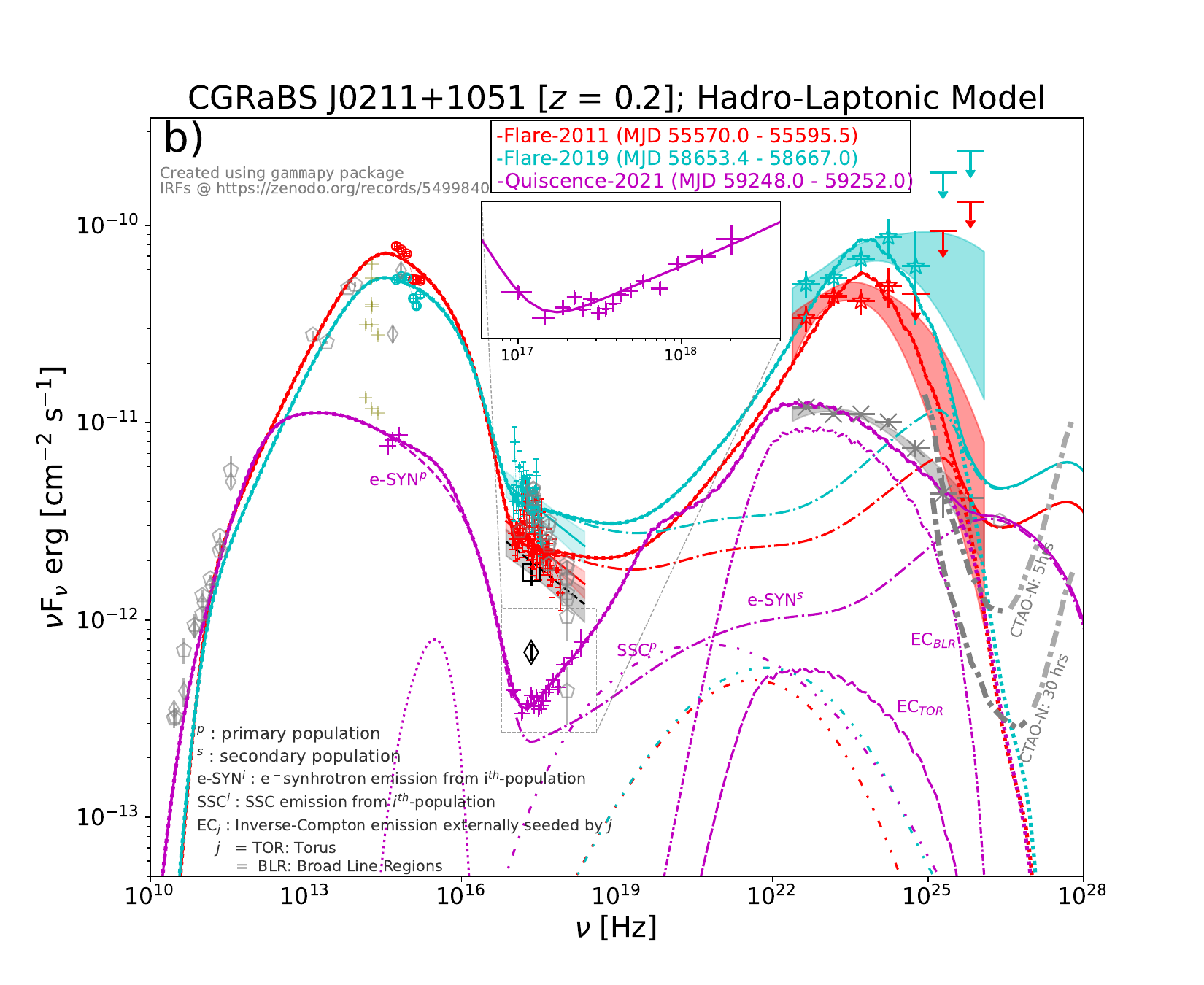}
    \includegraphics[width=0.35\textwidth]{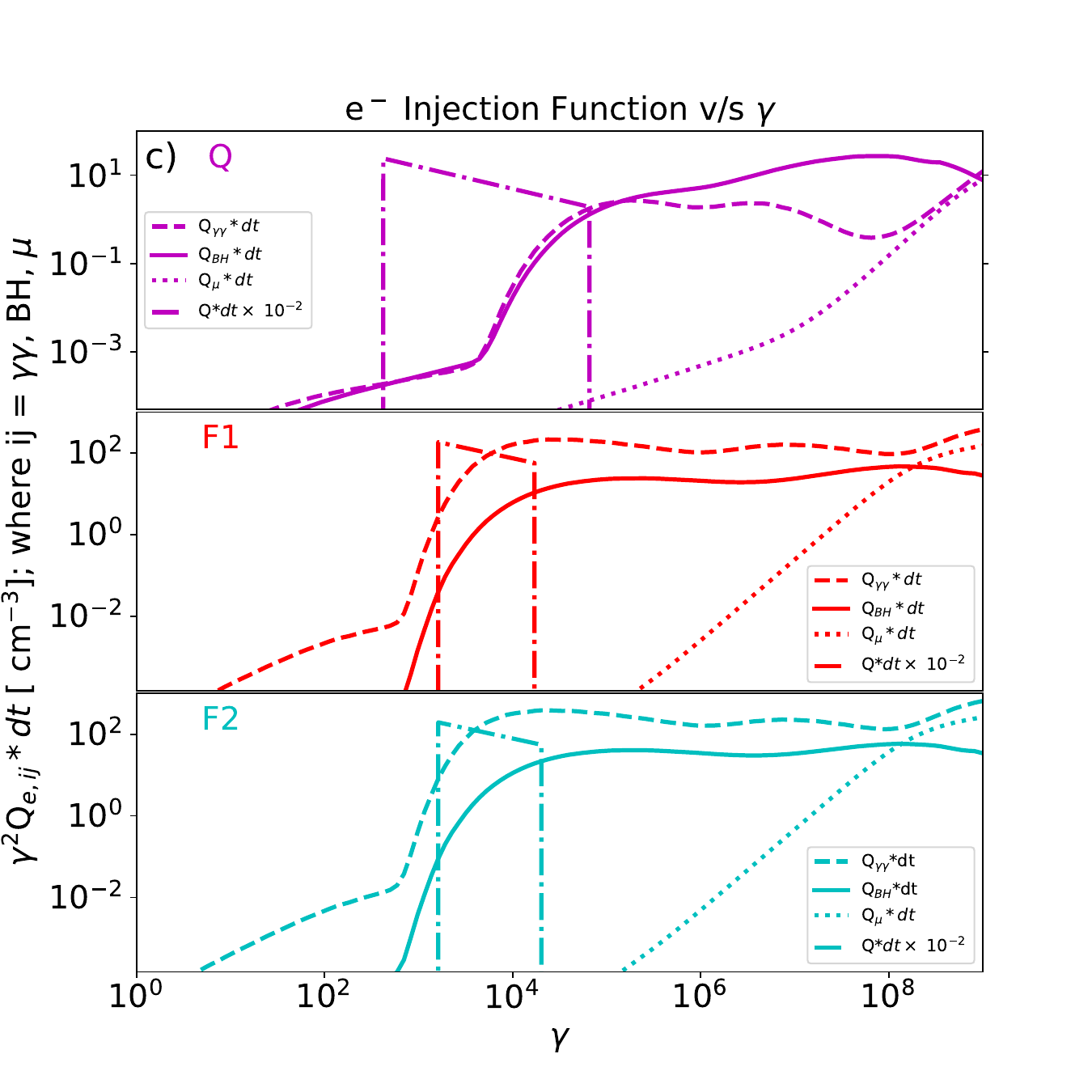}
    \includegraphics[width=0.42\textwidth]{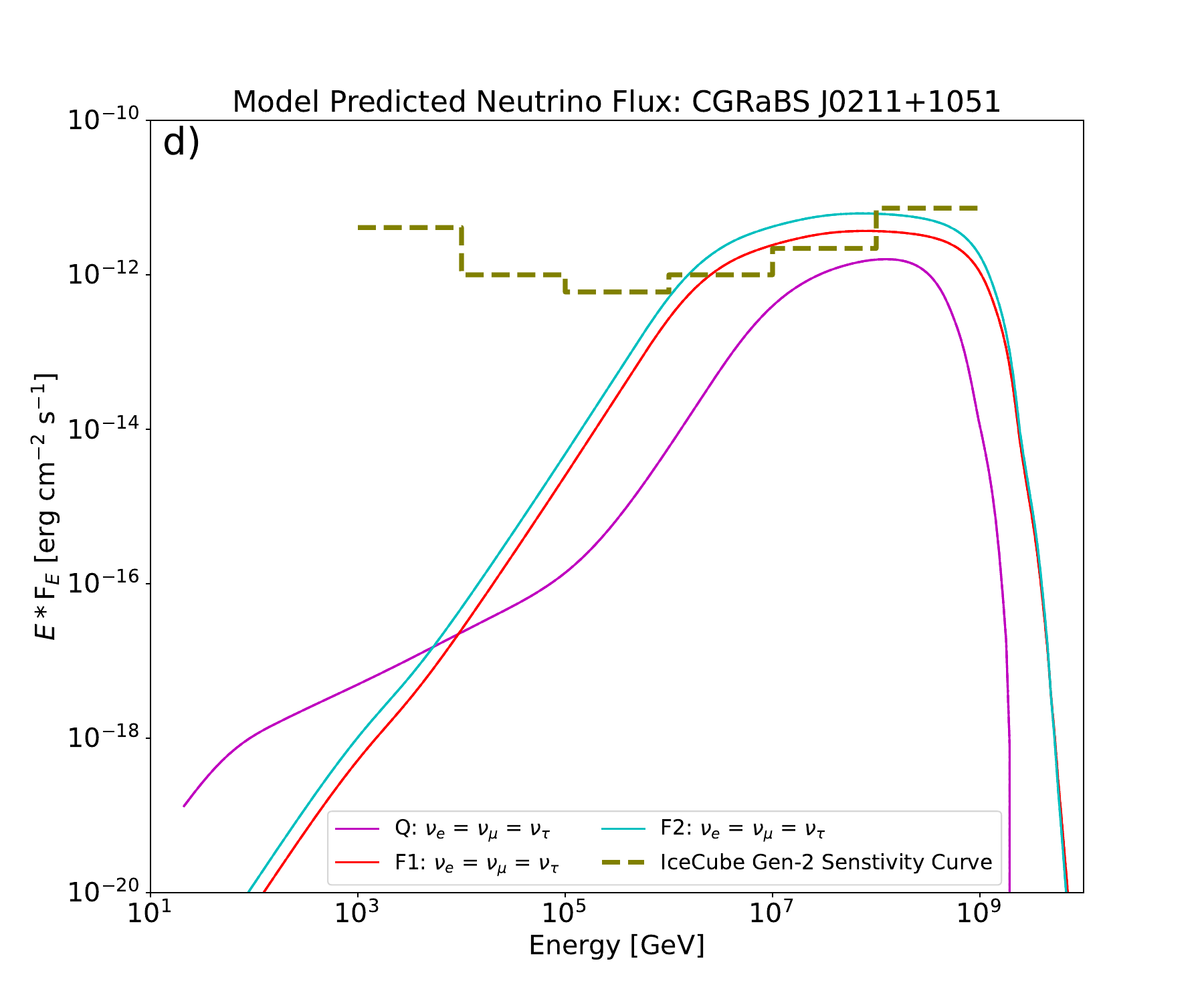}
    \caption{{\bf (a):} Multi-epoch SEDs (Q in magenta, F1 in red, and F2 in cyan) with the best-fit pure-leptonic model indicated by the correspondingly colored lines ({\bf solid: Total emission w/o EBL absorption, and, densely dotted extension of the same: with EBL absorption of  \citet{2017A&A...603A..34F}}). The shaded colored regions in $\gamma-$ray regime of the SEDs (E:$10^{22}-10^{26}$ Hz) of 12 year averaged (gray), F1 (red), and F2 (cyan), refer to the 1$\sigma$ uncertainty interval around the respective best fit {\tt LogParabola} models. {\bf (b):} Same as (a) but using a hadro-leptonic model. In both the SED plots, the dotted, dashed, dashed dotted, dashed double dotted, and densely dashed dotted magenta curves represent the BLR spectrum synchrotron, SSC, EC-BLR, and EC-DT components of Q, respectively. The inset shows the data and model for Q. These components are not plotted for F1 and F2 to keep the clarity in the figures. {\bf (c):} The injection of leptons of the various origins plotted with $\gamma$ (i.e., energy) derived from the solutions of the Fokker Plank Equation. {\bf (d):} The predicted neutrino ($\nu_\mu$,$\nu_e$,$\nu_\tau$) spectra corresponding to Q, F1, \& F2 for the hadro-leptonic model used in (b). The IceCube-Gen2 sensitivity is shown by the dashed curve in olive color.}
    \label{fig:mwsed-mdl}
\end{figure*}

A total of three different SEDs, as shown in Fig.~\ref{fig:mwsed-mdl}, were constructed and modeled using both leptonic and hadronic approaches to understand the emission processes. The first SED corresponds to the interval MJD 55570.0-55595.5 (F1: major overlap with F2011), the second to MJD 58653.4-58667.0 (F2: major overlap with F2019), and the third SED corresponds to MJD 59248-59252 (Q or Q2021), which covers the \xmm\ observations plus two ZTF observations before and after. All other historical data sets, mostly comprised of data taken before 2010, and not during the above epochs are also shown for the reference. For modeling the SED during Q, the 12-year averaged $\g-$ray SED, derived by our analysis, is used because the LAT analysis for this short duration results in mostly upper limits. These epochs identified by F2011, F2019, Q2021, etc., are clearly defined in Table \ref{tab:swiftdt}. 

 As demonstrated and argued in \S\ref{sec:uvxspecfit}, a simple power-law extrapolation of optical-UV fails to explain X-ray though a softer synchrotron-tail contribution cannot be ruled out. This suggests that the UV and X-ray emission is produced either by two separate radiation processes or both components originate from separate emission regions. The X-ray and $\gamma-$ray lightcurves observed during F1 and F2 reveal almost similar flaring behaviors, indicating co-spatial origin for the X-rays and $\gamma-$ray photons. We also notice significant flux enhancements in optical/UV bands during both flares. This strongly suggests that these outbursts originated from a single emission zone. Therefore, only a one-zone approach is considered in the modeling of the SEDs. However, we have used two different model setups, namely a pure-leptonic and a hadro-leptonic version. 

For the pure-leptonic case, we have modeled the broadband SED assuming that optical-UV is the synchrotron emission whereas the X-rays along with the $\gamma$ rays stem from the inverse-Comptonization of the seed photons in the emission region [i.e., SSC and EC as shown in Figure~\ref{fig:mwsed-mdl}-(a)]

The hadro-leptonic one-zone code \onehale\ \citep{Zacharias21,zacharias+22} is used to model these multi-epoch SEDs to constrain and investigate the physical parameters during the flares and quiescence. The \onehale\ model is a one-zone time-dependent code incorporating the evolution of both leptonic and hadronic primary and secondary particles (e$^-$,p$^+$, $\pi^0/\pi^-/\pi^+$, $\mu^-/\mu^+$) and their interactions resulting in radiative losses (synchrotron, inverse-Comptonization, and emission due to cascades through p-$\gamma$) in a spherical emission region self consistently.  
For example, the secondary electrons produced as the bi-product of hadronic cascades are added to the total electron density while considering the leptonic processes in every time step before reaching the equilibrium state or convergence. The photon density (n$_{ph}$) within the emission region in the code is evolved using the radiation transport equation \citep[Eq.~$\text{(3)}$ of][]{Zacharias21}. The evolution of the primary and secondary particles are evolved through the Fokker-Planck equation. 
Here, we only describe the purpose of the codes and the results, while a brief code description including definitions of the free parameters, is provided in the Appendix ~\ref{sec:codes}. In all cases, the SEDs have been derived as fits by eye. In such modeling efforts, a broad range of solutions is possible \citep[e.g.,][]{2020Galax...8...72C}. Table \ref{tab:sedparams} lists the important parameters for the employed models. Our strategy was to first reproduce the quiescence Q and then to modify a minimal set of input parameters to reproduce F1 and F2. 

\subsubsection{Pure-Leptonic Approach}
Since the low-energy hump is synchrotron radiation from relativistic electrons, inverse-Compton up-scattering of synchrotron photons is a natural process for the high-energy emission. Thus, the first attempt was to reproduce the high-energy part of the broadband SED using the SSC model only. 
 However, we were unable to find a suitable combination of parameters to fit the SEDs. Thus, we added the BLR and torus photon fields, which serve as additional target photons for the external-Compton (EC-IR/EC-BLR) process.
 
In our case, the BLR is defined by a mono-temperature black body (T=10000 K) in a spherical cell of radius 7.6$\times$10$^{16}$\,cm, and a total luminosity of 2.3$\times$ 10$^{41}$\,erg/s. The torus is parameterized by another mono-temperature black body (T=500 K), effective radius 4.2$\times$10$^{18}$\,cm, and total luminosity of 3.0$\times$10$^{41}$\,erg/s. The parameters characterizing the BLR and torus are typical to other BL Lac sources \citep[e.g., AP Librae,][]{zachariaswagner2016}. These input models of the BLR and torus are fixed throughout the study. 

As stated above, we first attempted to model Q and the data along with models  are shown in Fig. \ref{fig:mwsed-mdl}-(a). The broadband SED for Q is reasonably reproduced throughout the spectrum. However, for F1 and F2 only UV and $\gamma-$ray are reasonably reproduced whereas the X-ray spectra could not be reproduced. The reason is the observed soft UV spectra. The first three columns of the Table-\ref{tab:sedparams} summarize the best suitable parameters for this case. 

\subsubsection{Hadro-Leptonic Approach}

Following the failure of leptonic scenario for high states SED, the complete suits of the lepto-hadronic capability of the \onehale\ code were explored to reproduce the curious observed broadband emission features described above.

In this case, the excess X-ray emission is explained by synchrotron radiation from an additional ultra-relativistic lepton population produced as secondaries in hadronic interactions. It should be noted that the proton synchrotron process does not play any significant role in our model setup. 

Figure \ref{fig:mwsed-mdl}-(b) shows the best possible representation of the SED models. First, we derived the best-suited model parameters for Q. The turnover shape of Q is well explained using SSC with a minor contribution of emission from the secondary leptons resulting mainly from the $\gamma-\gamma$ interactions and from the cascades of the Bethe-Heitler mechanism. The evolution of the injection function for leptons of various origins, obtained from the solution of the Fokker–Planck equation, is shown in Figure-\ref{fig:mwsed-mdl}c. For the hadro-leptonic case, we had to use a larger $\gamma_{e,max}$ in comparison the to same for pure leptonic case to compensate for the turnover shape of X-ray SED during Q. 

Unlike in the pure-leptonic case, the broad-band SEDs of F1 and F2 are well explained using the hadro-leptonic model. This model also predicts a turnover in the hard X-ray band during both flares, which we could not observe possibly due to poor data.
                   
\section{Discussion and Summary}
\source has undergone two major $\gamma-$ray outbursts, the first in 2011 and the second in 2019. There was barely any activity in the \g-ray domain in the remaining time frame. On the contrary, the source is very active in optical and UV bands throughout the 12 year duration.

The major difference between the two outbursts, F1 and F2, is that the synchrotron peak flux during F2 is lower than during F1, whereas the \g\ peak flux is higher during F2 than F1. In short, the Compton ratio (ratio of the peak IC flux to peak synchrotron flux) is higher during F2 than F1. It is less than 1 during F1 larger than 1 during F2. The other peculiarity is the change is X-ray spectral shape during quiescent and bright phases of the source.
  
The multi-wavelength observations acquired over more than one decade indicate a softer UV spectrum than in the X-rays. The X-ray spectra during the deep quiescent state in 2021 clearly establish a turnover between 1.5-2 keV. Earlier modeling efforts by \citet{2022ApJ...931...59P} have also established the turnover using the same data. This is used to constrain the particle distribution responsible for the quiescent emission in 2021. The quiescent state Q is explained using the initial location of the emission region (d$_{loc}$) to be beyond the BLR region ($\ge$7.6 $\times$ 10$^{16}$ cm from the central engine). The initial location of the emission region (d$_{loc}$) is assigned a value such that it can optimally contribute to reproducing the observed GeV SED by boosting (within BLR) or de-boosting (outside the BLR) the target photon field for the EC. 

The softer spectral index in the UV, the observed X-ray emission, and high-energy $\gamma-$ray components are reproduced satisfactorily using the hadro-leptonic models. The size of emission regions derived from the models are of the same order or lower ($\sim$ 10$^{16}$cm) in comparison to the ones predicted by the variability time-scale from the flare modeling (ref. table \ref{tbl:flrfitmdl}). On the other hand, when we import GeV spectra from the second and third Catalog of Hard Fermi-LAT sources \citep{2016ApJS..222....5A, 2017ApJS..232...18A} and compare it with our 12-year averaged GeV SED, the composite $\gamma-$ray SED extends beyond 400 GeV. Our pure-leptonic model cannot explain such an extended $\gamma-$ray SED, whereas the hadro-leptonic model reproduces the $\gamma-$ray SED very well. This gives us confidence that the hadro-leptonic model has merits over pure-leptonic models in terms of explaining the broadband emissions from \source. 

The \onehale\ model clearly predicts a reasonably high flux in the TeV bands ($>$1 TeV), which may possibly be detected with upcoming, improved sensitivity VHE facility, the Cherenkov Telescope Array Observatory (CTAO\footnote{\url{https://www.ctao.org}}) even during its faint state. However, deeper X-ray observations are required during both flares and quiescence to test the full model predictions explored here. For example, the XMM observations during the quiescence state in 2021 revealed the spectral transition in Soft X-rays, which is very crucial information for constraining the particle spectrum. The curved UV-Xray SED may be a characteristic of \source\ even during flares, but the limited exposure does not allow us to derive strong conclusions during the flares. Our models predict a similar turn-over during both $\gamma-$ray flares. Further MWL monitoring would reveal the true nature of this source and potentially the TeV emission from ISPs.   
 
Our hadro-leptonic modeling also predicts potentially observable neutrino fluxes from \source\ as shown in Fig.~\ref{fig:mwsed-mdl}-(c).  
The model-predicted neutrino spectra are compared to the sensitivity of the IceCube Gen-2 adapted from \citet{2021JPhG...48f0501A}. Our modeling indicates that this source is not only a TeV emitter in the CTA era but also a neutrino candidate for IceCube Gen-2. It also strengthen the results on the possible correlation of 390 TeV IceCube neutrino events to radio flares in this object as indicated by \citep{2021A&A...650A..83H}. In short, this source has unique observational properties, such as a soft optical-UV spectra which does not connect straight to the X-rays. In \source\, the X-ray emission is possibly driven by the synchrotron emission of e$^+$-e$^-$ pairs, which are the bi-products of the hadronically-induced cascades. The predicted TeV ($\geq$1 TeV) and neutrino fluxes also make this source an important target in the multi-messenger era.   

\begin{table*}
\begin{center}
 {\footnotesize \caption{Best-fit SED model parameters for the three cases descried in \S \ref{sec:sedmod}\label{tab:sedparams}. 
  The bottom few rows are separated by a thick horizontal line referring to the estimated model output quantities.}
 \begin{tabular}{l|ccc|ccc}\hline
 \multirow{1}{*}{Name of the Parameters} &
 \multicolumn{3}{c|}{Pure Leptonic} &
 \multicolumn{3}{c}{Hadro-Leptonic} \\\hline
      & Q & F1 & F2 & Q  & F1 & F2\\ \hline
   Magnetic Field: B [Gauss]    & 0.065 & 0.5 & 0.6  & 0.085 & 0.75 & 0.55\\
   Size of Emission Region (R) [10$^{16}$ cm]  & 3.0 & 1.8 & 0.3 & 8.0 & 5.0 & 5.0\\
   Bulk Lorentz Factor: $\Gamma$ (= $\delta$)  & 30 & 30 & 35 & 25 & 27 & 27\\
   Min. Lorentz factor for $p$: $\gamma_{p, min}$ & -- & -- & --  & 50 & 1000 & 1000\\
   Max. Lorentz factor for $p$: $\gamma_{p, max}$ [10$^8$]  & -- & -- & -- & 3.0 & 7.0 & 7.0\\
   $p$ injection index: $s_p$  & -- & -- & -- & 2.1 & 2.1 & 2.1 \\
   $p$ Injection luminosity: L$_{p,j}$ [10$^{42}$ erg s$^{-1}$]  & -- & -- & -- & 22.5 & 6.5 & 10.0 \\
   Min. Lorentz factor for $e^-$: $\gamma_{e, min}$ & 300 & 1000 & 1000 & 400 & 1500 & 1500\\
   Max. Lorentz factor for $e^-$: $\gamma_{e, max}$ [10$^3$] & 80 & 8.0 & 8.0 & 70 & 20.0 & 23.0 \\
   Electron injection index: $s_e$  & 2.5 & 2.5 & 2.5 & 2.5 & 2.5 & 2.5 \\
   e$^-$ Injection luminosity: L$_{e,j}$ [10$^{41}$ erg s$^{-1}$] & 0.8 & 1.5 & 1.2 & 0.8 & 1.2 & 1.3 \\
   Initial location of the blob: d$_{loc}$ [10$^{16}$ cm] & 14.0 & 8.2 &  7.0 & 12.8 & 7.7 &  7.65 \\
   Mass of the blackhole: M$_{BH}$ [10$^8$ M$_\odot$]  & 2.4 & 2.4 & 2.4 & 2.4 & 2.4 & 2.4 \\ \hline
   
   Eddington luminosity: (L$_{edd}$) [10$^{46}$ erg s$^{-1}$]  & 3.02 & 3.02 &  3.02 & 3.02 & 3.02 &  3.02 \\ 
   Magnetic field luminosity: (L$_B$) [10$^{43}$ erg s$^{-1}$]  & 2.28 & 27.3 &  1.49  & 10.8 & 384.4 &  206.7 \\
   Proton luminosity: (L$_{p}$) [10$^{47}$ erg s$^{-1}$]  & -- & -- & -- & 18.6 & 0.7 &  1.04 \\
   Electron luminosity: (L$_{e}$) [10$^{44}$ erg s$^{-1}$]  & 8.77 & 3.84 & 3.71 & 4.23 & 0.44 & 0.54 \\
   Total radiative luminosity: (L$_{tot}$) [10$^{43}$ erg s$^{-1}$]  & 2.77 & 13.4 &  8.55  & 4.12 & 12.04 &  13.6 \\\hline
   Radius of the BLR region (R$_{BLR}$)[10$^{16}$ cm] & 7.6 & 7.6 & 7.6 & 7.6 & 7.6 & 7.6 \\
   Effective Luminosity of the BLR (L$_{BLR}$) [10$^{41}$ erg s$^{-1}$] & 8.0 & 8.0 & 8.0 & 8.0 & 8.0 & 8.0 \\
   Radius of the torus region (R$_{DT}$)[10$^{18}$ cm] & 4.2 & 4.2 & 4.2 & 4.2 & 4.2 & 4.2 \\
   Effective Luminosity of the torus (L$_{DT}$) [10$^{42}$ erg s$^{-1}$] & 3.0 & 3.0 & 3.0 & 3.0 & 3.0 & 3.0 \\
   \hline
 \end{tabular}}
\end{center}
 \end{table*}

Our modeling indicates that the emission from the X-ray to $\gamma$-ray bands in \source\ is significantly contributed by secondary leptons, highlighting the role of hadron-based interactions. The jet of \source\ therefore contains ultra-relativistic protons along with relativistic $e^-$ and $e^+$.  

\begin{acknowledgments}
S.C. acknowledges the supports from North-West University, South Africa and PRL Ahmedabad India to support this work. P.K. acknowledges support from the Department of Science and Technology (DST), Government of India, through the DST-INSPIRE Faculty grant (DST/INSPIRE/04/2020/002586). 
M.Z. acknowledges funding by the Deutsche Forschungsgemeinschaft (DFG, German Research Foundation) – project number 460248186 (PUNCH4NFDI).
This research has made use of data and/or software provided by the High Energy Astrophysics Science Archive Research Center (HEASARC), which is a service of the Astrophysics Science Division at NASA/GSFC and the High Energy Astrophysics Division of the Smithsonian Astrophysical Observatory. The authors also acknowledge the ZTF and CRTS team for publicly availing multi-filter optical observations.  
Data from the Steward Observatory spectropolarimetric monitoring project were used. This program is supported by Fermi Guest Investigator grants NNX08AW56G, NNX09AU10G, NNX12AO93G, and NNX15AU81G. 
This research has made use of public data from the ASDC SED Builder tool to access the publicly available data. This research has also made use of the NASA/IPAC Extragalactic Database (NED), which is operated by the Jet Propulsion Laboratory, California Institute of Technology, under contract with the National Aeronautics and Space Administration.  
\end{acknowledgments}
\vspace{5mm}
{\it Facilities:} Fermi (LAT), Swift, XMM-Newton, Steward Observatory, zTF, CRTS.\\

{\it Software:} HEASOFT (\url{https://heasarc.gsfc.nasa.gov/docs/software/heasoft/})

\bibliographystyle{aasjournal}
\bibliography{ms_biblifl}

@ARTICLE{Zacharias21,
       author = {{Zacharias}, Michael},
        title = "{Studying the Influence of External Photon Fields on Blazar Spectra Using a One-Zone Hadro-Leptonic Time-Dependent Model}",
      journal = {Physics},
         year = 2021,
        month = nov,
       volume = {3},
       number = {4},
        pages = {1098-1111},
          doi = {10.3390/physics3040069},
archivePrefix = {arXiv},
       eprint = {2111.05596},
 primaryClass = {astro-ph.HE},
       adsurl = {https://ui.adsabs.harvard.edu/abs/2021Physi...3.1098Z}
}

@ARTICLE{zacharias+22,
       author = {{Zacharias}, M. and {Reimer}, A. and {Boisson}, C. and {Zech}, A.},
        title = "{EXHALE-JET: an extended hadro-leptonic jet model for blazars - I. Code description and initial results}",
      journal = {\mnras},
         year = 2022,
        month = may,
       volume = {512},
       number = {3},
        pages = {3948-3971},
          doi = {10.1093/mnras/stac754},
archivePrefix = {arXiv},
       eprint = {2203.07956},
 primaryClass = {astro-ph.HE},
       adsurl = {https://ui.adsabs.harvard.edu/abs/2022MNRAS.512.3948Z}
}

@ARTICLE{2016ApJS..222....5A,
       author = {{Ackermann}, M. and {Ajello}, M. and {Atwood}, W.~B. and {Baldini}, L. and {Ballet}, J. and {Barbiellini}, G. and {Bastieri}, D. and {Becerra Gonzalez}, J. and {Bellazzini}, R. and {Bissaldi}, E. and {Blandford}, R.~D. and {Bloom}, E.~D. and {Bonino}, R. and {Bottacini}, E. and {Brandt}, T.~J. and {Bregeon}, J. and {Bruel}, P. and {Buehler}, R. and {Buson}, S. and {Caliandro}, G.~A. and {Cameron}, R.~A. and {Caputo}, R. and {Caragiulo}, M. and {Caraveo}, P.~A. and {Cavazzuti}, E. and {Cecchi}, C. and {Charles}, E. and {Chekhtman}, A. and {Cheung}, C.~C. and {Chiang}, J. and {Chiaro}, G. and {Ciprini}, S. and {Cohen}, J.~M. and {Cohen-Tanugi}, J. and {Cominsky}, L.~R. and {Conrad}, J. and {Cuoco}, A. and {Cutini}, S. and {D'Ammando}, F. and {de Angelis}, A. and {de Palma}, F. and {Desiante}, R. and {Di Mauro}, M. and {Di Venere}, L. and {Dom{\'\i}nguez}, A. and {Drell}, P.~S. and {Favuzzi}, C. and {Fegan}, S.~J. and {Ferrara}, E.~C. and {Focke}, W.~B. and {Fortin}, P. and {Franckowiak}, A. and {Fukazawa}, Y. and {Funk}, S. and {Furniss}, A.~K. and {Fusco}, P. and {Gargano}, F. and {Gasparrini}, D. and {Giglietto}, N. and {Giommi}, P. and {Giordano}, F. and {Giroletti}, M. and {Glanzman}, T. and {Godfrey}, G. and {Grenier}, I.~A. and {Grondin}, M. -H. and {Guillemot}, L. and {Guiriec}, S. and {Harding}, A.~K. and {Hays}, E. and {Hewitt}, J.~W. and {Hill}, A.~B. and {Horan}, D. and {Iafrate}, G. and {Hartmann}, Dieter and {Jogler}, T. and {J{\'o}hannesson}, G. and {Johnson}, A.~S. and {Kamae}, T. and {Kataoka}, J. and {Kn{\"o}dlseder}, J. and {Kuss}, M. and {La Mura}, G. and {Larsson}, S. and {Latronico}, L. and {Lemoine-Goumard}, M. and {Li}, J. and {Li}, L. and {Longo}, F. and {Loparco}, F. and {Lott}, B. and {Lovellette}, M.~N. and {Lubrano}, P. and {Madejski}, G.~M. and {Maldera}, S. and {Manfreda}, A. and {Mayer}, M. and {Mazziotta}, M.~N. and {Michelson}, P.~F. and {Mirabal}, N. and {Mitthumsiri}, W. and {Mizuno}, T. and {Moiseev}, A.~A. and {Monzani}, M.~E. and {Morselli}, A. and {Moskalenko}, I.~V. and {Murgia}, S. and {Nuss}, E. and {Ohsugi}, T. and {Omodei}, N. and {Orienti}, M. and {Orlando}, E. and {Ormes}, J.~F. and {Paneque}, D. and {Perkins}, J.~S. and {Pesce-Rollins}, M. and {Petrosian}, V. and {Piron}, F. and {Pivato}, G. and {Porter}, T.~A. and {Rain{\`o}}, S. and {Rando}, R. and {Razzano}, M. and {Razzaque}, S. and {Reimer}, A. and {Reimer}, O. and {Reposeur}, T. and {Romani}, R.~W. and {S{\'a}nchez-Conde}, M. and {Saz Parkinson}, P.~M. and {Schmid}, J. and {Schulz}, A. and {Sgr{\`o}}, C. and {Siskind}, E.~J. and {Spada}, F. and {Spandre}, G. and {Spinelli}, P. and {Suson}, D.~J. and {Tajima}, H. and {Takahashi}, H. and {Takahashi}, M. and {Takahashi}, T. and {Thayer}, J.~B. and {Thompson}, D.~J. and {Tibaldo}, L. and {Torres}, D.~F. and {Tosti}, G. and {Troja}, E. and {Vianello}, G. and {Wood}, K.~S. and {Wood}, M. and {Yassine}, M. and {Zaharijas}, G. and {Zimmer}, S.},
        title = "{2FHL: The Second Catalog of Hard Fermi-LAT Sources}",
      journal = {\apjs},
         year = 2016,
        month = jan,
       volume = {222},
       number = {1},
          eid = {5},
        pages = {5},
          doi = {10.3847/0067-0049/222/1/5},
archivePrefix = {arXiv},
       eprint = {1508.04449},
 primaryClass = {astro-ph.HE},
       adsurl = {https://ui.adsabs.harvard.edu/abs/2016ApJS..222....5A}
}

@ARTICLE{2017ApJS..232...18A,
       author = {{Ajello}, M. and {Atwood}, W.~B. and {Baldini}, L. and {Ballet}, J. and {Barbiellini}, G. and {Bastieri}, D. and {Bellazzini}, R. and {Bissaldi}, E. and {Blandford}, R.~D. and {Bloom}, E.~D. and {Bonino}, R. and {Bregeon}, J. and {Britto}, R.~J. and {Bruel}, P. and {Buehler}, R. and {Buson}, S. and {Cameron}, R.~A. and {Caputo}, R. and {Caragiulo}, M. and {Caraveo}, P.~A. and {Cavazzuti}, E. and {Cecchi}, C. and {Charles}, E. and {Chekhtman}, A. and {Cheung}, C.~C. and {Chiaro}, G. and {Ciprini}, S. and {Cohen}, J.~M. and {Costantin}, D. and {Costanza}, F. and {Cuoco}, A. and {Cutini}, S. and {D'Ammando}, F. and {de Palma}, F. and {Desiante}, R. and {Digel}, S.~W. and {Di Lalla}, N. and {Di Mauro}, M. and {Di Venere}, L. and {Dom{\'\i}nguez}, A. and {Drell}, P.~S. and {Dumora}, D. and {Favuzzi}, C. and {Fegan}, S.~J. and {Ferrara}, E.~C. and {Fortin}, P. and {Franckowiak}, A. and {Fukazawa}, Y. and {Funk}, S. and {Fusco}, P. and {Gargano}, F. and {Gasparrini}, D. and {Giglietto}, N. and {Giommi}, P. and {Giordano}, F. and {Giroletti}, M. and {Glanzman}, T. and {Green}, D. and {Grenier}, I.~A. and {Grondin}, M. -H. and {Grove}, J.~E. and {Guillemot}, L. and {Guiriec}, S. and {Harding}, A.~K. and {Hays}, E. and {Hewitt}, J.~W. and {Horan}, D. and {J{\'o}hannesson}, G. and {Kensei}, S. and {Kuss}, M. and {La Mura}, G. and {Larsson}, S. and {Latronico}, L. and {Lemoine-Goumard}, M. and {Li}, J. and {Longo}, F. and {Loparco}, F. and {Lott}, B. and {Lubrano}, P. and {Magill}, J.~D. and {Maldera}, S. and {Manfreda}, A. and {Mazziotta}, M.~N. and {McEnery}, J.~E. and {Meyer}, M. and {Michelson}, P.~F. and {Mirabal}, N. and {Mitthumsiri}, W. and {Mizuno}, T. and {Moiseev}, A.~A. and {Monzani}, M.~E. and {Morselli}, A. and {Moskalenko}, I.~V. and {Negro}, M. and {Nuss}, E. and {Ohsugi}, T. and {Omodei}, N. and {Orienti}, M. and {Orlando}, E. and {Palatiello}, M. and {Paliya}, V.~S. and {Paneque}, D. and {Perkins}, J.~S. and {Persic}, M. and {Pesce-Rollins}, M. and {Piron}, F. and {Porter}, T.~A. and {Principe}, G. and {Rain{\`o}}, S. and {Rando}, R. and {Razzano}, M. and {Razzaque}, S. and {Reimer}, A. and {Reimer}, O. and {Reposeur}, T. and {Saz Parkinson}, P.~M. and {Sgr{\`o}}, C. and {Simone}, D. and {Siskind}, E.~J. and {Spada}, F. and {Spandre}, G. and {Spinelli}, P. and {Stawarz}, L. and {Suson}, D.~J. and {Takahashi}, M. and {Tak}, D. and {Thayer}, J.~G. and {Thayer}, J.~B. and {Thompson}, D.~J. and {Torres}, D.~F. and {Torresi}, E. and {Troja}, E. and {Vianello}, G. and {Wood}, K. and {Wood}, M.},
        title = "{3FHL: The Third Catalog of Hard Fermi-LAT Sources}",
      journal = {\apjs},
         year = 2017,
        month = oct,
       volume = {232},
       number = {2},
          eid = {18},
        pages = {18},
          doi = {10.3847/1538-4365/aa8221},
archivePrefix = {arXiv},
       eprint = {1702.00664},
 primaryClass = {astro-ph.HE},
       adsurl = {https://ui.adsabs.harvard.edu/abs/2017ApJS..232...18A}
}

@ARTICLE{2015MNRAS.450.2658M,
       author = {{Mufakharov}, T. and {Mingaliev}, M. and {Sotnikova}, Yu. and {Naiden}, Ya. and {Erkenov}, A.},
        title = "{The observed radio/gamma-ray emission correlation for blazars with the Fermi-LAT and the RATAN-600 data}",
      journal = {\mnras},
         year = 2015,
        month = jul,
       volume = {450},
       number = {3},
        pages = {2658-2669},
          doi = {10.1093/mnras/stv772},
archivePrefix = {arXiv},
       eprint = {1504.06704},
 primaryClass = {astro-ph.HE},
       adsurl = {https://ui.adsabs.harvard.edu/abs/2015MNRAS.450.2658M}
}

@article{Abdo_2010,
doi = {10.1088/0067-0049/188/2/405},
url = {https://dx.doi.org/10.1088/0067-0049/188/2/405},
year = {2010},
month = {may},
publisher = {The American Astronomical Society},
volume = {188},
number = {2},
pages = {405},
author = {A. A. Abdo and M. Ackermann and M. Ajello and A. Allafort and E. Antolini and W. B. Atwood and M. Axelsson and L. Baldini and J. Ballet and G. Barbiellini and D. Bastieri and B. M. Baughman and K. Bechtol and R. Bellazzini and F. Belli and B. Berenji and D. Bisello and R. D. Blandford and E. D. Bloom and E. Bonamente and J. Bonnell and A. W. Borgland and A. Bouvier and J. Bregeon and A. Brez and M. Brigida and P. Bruel and T. H. Burnett and G. Busetto and S. Buson and G. A. Caliandro and R. A. Cameron and R. Campana and B. Canadas and P. A. Caraveo and S. Carrigan and J. M. Casandjian and E. Cavazzuti and M. Ceccanti and C. Cecchi and Ö. Çelik and E. Charles and A. Chekhtman and C. C. Cheung and J. Chiang and A. N. Cillis and S. Ciprini and R. Claus and J. Cohen-Tanugi and J. Conrad and R. Corbet and D. S. Davis and M. DeKlotz and P. R. den Hartog and C. D. Dermer and A. de Angelis and A. de Luca and F. de Palma and S. W. Digel and M. Dormody and E. do Couto e Silva and P. S. Drell and R. Dubois and D. Dumora and D. Fabiani and C. Farnier and C. Favuzzi and S. J. Fegan and E. C. Ferrara and W. B. Focke and P. Fortin and M. Frailis and Y. Fukazawa and S. Funk and P. Fusco and F. Gargano and D. Gasparrini and N. Gehrels and S. Germani and G. Giavitto and B. Giebels and N. Giglietto and P. Giommi and F. Giordano and M. Giroletti and T. Glanzman and G. Godfrey and I. A. Grenier and M.-H. Grondin and J. E. Grove and L. Guillemot and S. Guiriec and M. Gustafsson and D. Hadasch and Y. Hanabata and A. K. Harding and M. Hayashida and E. Hays and S. E. Healey and A. B. Hill and D. Horan and R. E. Hughes and G. Iafrate and G. Jóhannesson and A. S. Johnson and R. P. Johnson and T. J. Johnson and W. N. Johnson and T. Kamae and H. Katagiri and J. Kataoka and N. Kawai and M. Kerr and J. Knödlseder and D. Kocevski and M. Kuss and J. Lande and D. Landriu and L. Latronico and S.-H. Lee and M. Lemoine-Goumard and A. M. Lionetto and M. Llena Garde and F. Longo and F. Loparco and B. Lott and M. N. Lovellette and P. Lubrano and G. M. Madejski and A. Makeev and B. Marangelli and M. Marelli and E. Massaro and M. N. Mazziotta and W. McConville and J. E. McEnery and P. F. Michelson and M. Minuti and W. Mitthumsiri and T. Mizuno and A. A. Moiseev and M. Mongelli and C. Monte and M. E. Monzani and E. Moretti and A. Morselli and I. V. Moskalenko and S. Murgia and H. Nakajima and T. Nakamori and M. Naumann-Godo and P. L. Nolan and J. P. Norris and E. Nuss and M. Ohno and T. Ohsugi and N. Omodei and E. Orlando and J. F. Ormes and M. Ozaki and A. Paccagnella and D. Paneque and J. H. Panetta and D. Parent and V. Pelassa and M. Pepe and M. Pesce-Rollins and M. Pinchera and F. Piron and T. A. Porter and L. Poupard and S. Rainò and R. Rando and P. S. Ray and M. Razzano and S. Razzaque and N. Rea and A. Reimer and O. Reimer and T. Reposeur and J. Ripken and S. Ritz and L. S. Rochester and A. Y. Rodriguez and R. W. Romani and M. Roth and H. F.-W. Sadrozinski and D. Salvetti and D. Sanchez and A. Sander and P. M. Saz Parkinson and J. D. Scargle and T. L. Schalk and G. Scolieri and C. Sgrò and M. S. Shaw and E. J. Siskind and D. A. Smith and P. D. Smith and G. Spandre and P. Spinelli and J.-L. Starck and T. E. Stephens and E. Striani and M. S. Strickman and A. W. Strong and D. J. Suson and H. Tajima and H. Takahashi and T. Takahashi and T. Tanaka and J. B. Thayer and J. G. Thayer and D. J. Thompson and L. Tibaldo and O. Tibolla and F. Tinebra and D. F. Torres and G. Tosti and A. Tramacere and Y. Uchiyama and T. L. Usher and A. Van Etten and V. Vasileiou and N. Vilchez and V. Vitale and A. P. Waite and E. Wallace and P. Wang and K. Watters and B. L. Winer and K. S. Wood and Z. Yang and T. Ylinen and M. Ziegler},
title = {FERMI LARGE AREA TELESCOPE FIRST SOURCE
                    CATALOG},
journal = {The Astrophysical Journal Supplement Series}
}

@ARTICLE{2010ApJ...722..520A,
       author = {{Abdo}, A.~A. and {Ackermann}, M. and {Ajello}, M. and {Antolini}, E. and {Baldini}, L. and {Ballet}, J. and {Barbiellini}, G. and {Bastieri}, D. and {Bechtol}, K. and {Bellazzini}, R. and {Berenji}, B. and {Blandford}, R.~D. and {Bloom}, E.~D. and {Bonamente}, E. and {Borgland}, A.~W. and {Bouvier}, A. and {Bregeon}, J. and {Brez}, A. and {Brigida}, M. and {Bruel}, P. and {Buehler}, R. and {Burnett}, T.~H. and {Buson}, S. and {Caliandro}, G.~A. and {Cameron}, R.~A. and {Caraveo}, P.~A. and {Carrigan}, S. and {Casandjian}, J.~M. and {Cavazzuti}, E. and {Cecchi}, C. and {{\c{C}}elik}, {\"O}. and {Chekhtman}, A. and {Cheung}, C.~C. and {Chiang}, J. and {Ciprini}, S. and {Claus}, R. and {Cohen-Tanugi}, J. and {Cominsky}, L.~R. and {Conrad}, J. and {Costamante}, L. and {Cutini}, S. and {Dermer}, C.~D. and {de Angelis}, A. and {de Palma}, F. and {Silva}, E. do Couto e. and {Drell}, P.~S. and {Dubois}, R. and {Dumora}, D. and {Farnier}, C. and {Favuzzi}, C. and {Fegan}, S.~J. and {Focke}, W.~B. and {Fortin}, P. and {Frailis}, M. and {Fukazawa}, Y. and {Funk}, S. and {Fusco}, P. and {Gargano}, F. and {Gasparrini}, D. and {Gehrels}, N. and {Germani}, S. and {Giebels}, B. and {Giglietto}, N. and {Giommi}, P. and {Giordano}, F. and {Glanzman}, T. and {Godfrey}, G. and {Grenier}, I.~A. and {Grondin}, M. -H. and {Grove}, J.~E. and {Guiriec}, S. and {Hadasch}, D. and {Hayashida}, M. and {Hays}, E. and {Healey}, S.~E. and {Horan}, D. and {Hughes}, R.~E. and {Itoh}, R. and {J{\'o}hannesson}, G. and {Johnson}, A.~S. and {Johnson}, W.~N. and {Kamae}, T. and {Katagiri}, H. and {Kataoka}, J. and {Kawai}, N. and {Kn{\"o}dlseder}, J. and {Kuss}, M. and {Lande}, J. and {Larsson}, S. and {Latronico}, L. and {Lemoine-Goumard}, M. and {Longo}, F. and {Loparco}, F. and {Lott}, B. and {Lovellette}, M.~N. and {Lubrano}, P. and {Madejski}, G.~M. and {Makeev}, A. and {Massaro}, E. and {Mazziotta}, M.~N. and {McEnery}, J.~E. and {Michelson}, P.~F. and {Mitthumsiri}, W. and {Mizuno}, T. and {Moiseev}, A.~A. and {Monte}, C. and {Monzani}, M.~E. and {Morselli}, A. and {Moskalenko}, I.~V. and {Mueller}, M. and {Murgia}, S. and {Nolan}, P.~L. and {Norris}, J.~P. and {Nuss}, E. and {Ohno}, M. and {Ohsugi}, T. and {Omodei}, N. and {Orlando}, E. and {Ormes}, J.~F. and {Ozaki}, M. and {Panetta}, J.~H. and {Parent}, D. and {Pelassa}, V. and {Pepe}, M. and {Pesce-Rollins}, M. and {Piron}, F. and {Porter}, T.~A. and {Rain{\`o}}, S. and {Rando}, R. and {Razzano}, M. and {Reimer}, A. and {Reimer}, O. and {Ritz}, S. and {Rodriguez}, A.~Y. and {Romani}, R.~W. and {Roth}, M. and {Ryde}, F. and {Sadrozinski}, H.~F. -W. and {Sander}, A. and {Scargle}, J.~D. and {Sgr{\`o}}, C. and {Shaw}, M.~S. and {Smith}, P.~D. and {Spandre}, G. and {Spinelli}, P. and {Starck}, J. -L. and {Strickman}, M.~S. and {Suson}, D.~J. and {Takahashi}, H. and {Takahashi}, T. and {Tanaka}, T. and {Thayer}, J.~B. and {Thayer}, J.~G. and {Thompson}, D.~J. and {Tibaldo}, L. and {Torres}, D.~F. and {Tosti}, G. and {Tramacere}, A. and {Uchiyama}, Y. and {Usher}, T.~L. and {Vasileiou}, V. and {Vilchez}, N. and {Vitale}, V. and {Waite}, A.~P. and {Wallace}, E. and {Wang}, P. and {Winer}, B.~L. and {Wood}, K.~S. and {Yang}, Z. and {Ylinen}, T. and {Ziegler}, M.},
        title = "{Gamma-ray Light Curves and Variability of Bright Fermi-detected Blazars}",
      journal = {\apj},
         year = 2010,
        month = oct,
       volume = {722},
       number = {1},
        pages = {520-542},
          doi = {10.1088/0004-637X/722/1/520},
archivePrefix = {arXiv},
       eprint = {1004.0348},
 primaryClass = {astro-ph.HE},
       adsurl = {https://ui.adsabs.harvard.edu/abs/2010ApJ...722..520A}
}

@article{Hummer_2010,
doi = {10.1088/0004-637X/721/1/630},
url = {https://dx.doi.org/10.1088/0004-637X/721/1/630},
year = {2010},
month = {aug},
publisher = {The American Astronomical Society},
volume = {721},
number = {1},
pages = {630},
author = {S. Hümmer and M. Rüger and F. Spanier and W. Winter},
title = {SIMPLIFIED MODELS FOR PHOTOHADRONIC INTERACTIONS IN COSMIC ACCELERATORS},
journal = {The Astrophysical Journal},
}

@ARTICLE{1982ApJ...253...38U,
       author = {{Urry}, C.~M. and {Mushotzky}, R.~F.},
        title = "{PKS 2155-304 : relativistically beamed synchrotron radiation from a BL Lacertae object.}",
      journal = {\apj},
         year = 1982,
        month = feb,
       volume = {253},
        pages = {38-46},
          doi = {10.1086/159607},
       adsurl = {https://ui.adsabs.harvard.edu/abs/1982ApJ...253...38U}
}

@ARTICLE{1989MNRAS.236..341G,
       author = {{Ghisellini}, Gabriele},
        title = "{Synchrotron self Compton models for compact sources - The case of a steep power-law particle distribution}",
      journal = {\mnras},
         year = 1989,
        month = jan,
       volume = {236},
        pages = {341-351},
          doi = {10.1093/mnras/236.2.341},
       adsurl = {https://ui.adsabs.harvard.edu/abs/1989MNRAS.236..341G}
}

@ARTICLE{2017A&A...603A..34F,
       author = {{Franceschini}, Alberto and {Rodighiero}, Giulia},
        title = "{The extragalactic background light revisited and the cosmic photon-photon opacity}",
      journal = {\aap},
         year = 2017,
        month = jul,
       volume = {603},
          eid = {A34},
        pages = {A34},
          doi = {10.1051/0004-6361/201629684},
archivePrefix = {arXiv},
       eprint = {1705.10256},
 primaryClass = {astro-ph.HE},
       adsurl = {https://ui.adsabs.harvard.edu/abs/2017A&A...603A..34F}
}

@ARTICLE{2021A&A...650A..83H,
       author = {{Hovatta}, T. and {Lindfors}, E. and {Kiehlmann}, S. and {Max-Moerbeck}, W. and {Hodges}, M. and {Liodakis}, I. and {L{\"a}hteem{\"a}ki}, A. and {Pearson}, T.~J. and {Readhead}, A.~C.~S. and {Reeves}, R.~A. and {Suutarinen}, S. and {Tammi}, J. and {Tornikoski}, M.},
        title = "{Association of IceCube neutrinos with radio sources observed at Owens Valley and Mets{\"a}hovi Radio Observatories}",
      journal = {\aap},
         year = 2021,
        month = jun,
       volume = {650},
          eid = {A83},
        pages = {A83},
          doi = {10.1051/0004-6361/202039481},
archivePrefix = {arXiv},
       eprint = {2009.10523},
 primaryClass = {astro-ph.HE},
       adsurl = {https://ui.adsabs.harvard.edu/abs/2021A&A...650A..83H}
}

@ARTICLE{2021JPhG...48f0501A,
       author = {{Aartsen}, M.~G. and {Abbasi}, R. and {Ackermann}, M. and {Adams}, J. and {Aguilar}, J.~A. and {Ahlers}, M. and {Ahrens}, M. and {Alispach}, C. and {Allison}, P. and {Amin}, N.~M. and {Andeen}, K. and {Anderson}, T. and {Ansseau}, I. and {Anton}, G. and {Arg{\"u}elles}, C. and {Arlen}, T.~C. and {Auffenberg}, J. and {Axani}, S. and {Bagherpour}, H. and {Bai}, X. and {Balagopal V}, A. and {Barbano}, A. and {Bartos}, I. and {Bastian}, B. and {Basu}, V. and {Baum}, V. and {Baur}, S. and {Bay}, R. and {Beatty}, J.~J. and {Becker}, K. -H. and {Tjus}, J. Becker and {BenZvi}, S. and {Berley}, D. and {Bernardini}, E. and {Besson}, D.~Z. and {Binder}, G. and {Bindig}, D. and {Blaufuss}, E. and {Blot}, S. and {Bohm}, C. and {Bohmer}, M. and {B{\"o}ser}, S. and {Botner}, O. and {B{\"o}ttcher}, J. and {Bourbeau}, E. and {Bourbeau}, J. and {Bradascio}, F. and {Braun}, J. and {Bron}, S. and {Brostean-Kaiser}, J. and {Burgman}, A. and {Burley}, R.~T. and {Buscher}, J. and {Busse}, R.~S. and {Bustamante}, M. and {Campana}, M.~A. and {Carnie-Bronca}, E.~G. and {Carver}, T. and {Chen}, C. and {Chen}, P. and {Cheung}, E. and {Chirkin}, D. and {Choi}, S. and {Clark}, B.~A. and {Clark}, K. and {Classen}, L. and {Coleman}, A. and {Collin}, G.~H. and {Connolly}, A. and {Conrad}, J.~M. and {Coppin}, P. and {Correa}, P. and {Cowen}, D.~F. and {Cross}, R. and {Dave}, P. and {Deaconu}, C. and {De Clercq}, C. and {DeLaunay}, J.~J. and {De Kockere}, S. and {Dembinski}, H. and {Deoskar}, K. and {De Ridder}, S. and {Desai}, A. and {Desiati}, P. and {de Vries}, K.~D. and {de Wasseige}, G. and {de With}, M. and {DeYoung}, T. and {Dharani}, S. and {Diaz}, A. and {D{\'\i}az-V{\'e}lez}, J.~C. and {Dujmovic}, H. and {Dunkman}, M. and {DuVernois}, M.~A. and {Dvorak}, E. and {Ehrhardt}, T. and {Eller}, P. and {Engel}, R. and {Evans}, J.~J. and {Evenson}, P.~A. and {Fahey}, S. and {Farrag}, K. and {Fazely}, A.~R. and {Felde}, J. and {Fienberg}, A.~T. and {Filimonov}, K. and {Finley}, C. and {Fischer}, L. and {Fox}, D. and {Franckowiak}, A. and {Friedman}, E. and {Fritz}, A. and {Gaisser}, T.~K. and {Gallagher}, J. and {Ganster}, E. and {Garcia-Fernandez}, D. and {Garrappa}, S. and {Gartner}, A. and {Gerhard}, L. and {Gernhaeuser}, R. and {Ghadimi}, A. and {Glaser}, C. and {Glauch}, T. and {Gl{\"u}senkamp}, T. and {Goldschmidt}, A. and {Gonzalez}, J.~G. and {Goswami}, S. and {Grant}, D. and {Gr{\'e}goire}, T. and {Griffith}, Z. and {Griswold}, S. and {G{\"u}nd{\"u}z}, M. and {Haack}, C. and {Hallgren}, A. and {Halliday}, R. and {Halve}, L. and {Halzen}, F. and {Hanson}, J.~C. and {Hanson}, K. and {Hardin}, J. and {Haugen}, J. and {Haungs}, A. and {Hauser}, S. and {Hebecker}, D. and {Heinen}, D. and {Heix}, P. and {Helbing}, K. and {Hellauer}, R. and {Henningsen}, F. and {Hickford}, S. and {Hignight}, J. and {Hill}, C. and {Hill}, G.~C. and {Hoffman}, K.~D. and {Hoffmann}, B. and {Hoffmann}, R. and {Hoinka}, T. and {Hokanson-Fasig}, B. and {Holzapfel}, K. and {Hoshina}, K. and {Huang}, F. and {Huber}, M. and {Huber}, T. and {Huege}, T. and {Hughes}, K. and {Hultqvist}, K. and {H{\"u}nnefeld}, M. and {Hussain}, R. and {In}, S. and {Iovine}, N. and {Ishihara}, A. and {Jansson}, M. and {Japaridze}, G.~S. and {Jeong}, M. and {Jones}, B.~J.~P. and {Jonske}, F. and {Joppe}, R. and {Kalekin}, O. and {Kang}, D. and {Kang}, W. and {Kang}, X. and {Kappes}, A. and {Kappesser}, D. and {Karg}, T. and {Karl}, M. and {Karle}, A. and {Katori}, T. and {Katz}, U. and {Kauer}, M. and {Keivani}, A. and {Kellermann}, M. and {Kelley}, J.~L. and {Kheirandish}, A. and {Kim}, J. and {Kin}, K. and {Kintscher}, T. and {Kiryluk}, J. and {Kittler}, T. and {Kleifges}, M. and {Klein}, S.~R. and {Koirala}, R. and {Kolanoski}, H. and {K{\"o}pke}, L. and {Kopper}, C. and {Kopper}, S. and {Koskinen}, D.~J. and {Koundal}, P. and {Kovacevich}, M. and {Kowalski}, M. and {Krauss}, C.~B. and {Krings}, K. and {Kr{\"u}ckl}, G. and {Kulacz}, N. and {Kurahashi}, N. and {Gualda}, C. Lagunas and {Lahmann}, R. and {Lanfranchi}, J.~L. and {Larson}, M.~J. and {Latif}, U. and {Lauber}, F. and {Lazar}, J.~P. and {Leonard}, K. and {Leszczy{\'n}ska}, A. and {Li}, Y. and {Liu}, Q.~R. and {Lohfink}, E. and {LoSecco}, J. and {Mariscal}, C.~J. Lozano and {Lu}, L. and {Lucarelli}, F. and {Ludwig}, A. and {L{\"u}nemann}, J. and {Luszczak}, W. and {Lyu}, Y. and {Ma}, W.~Y. and {Madsen}, J. and {Maggi}, G. and {Mahn}, K.~B.~M. and {Makino}, Y. and {Mallik}, P. and {Mancina}, S. and {Mandalia}, S. and {Mari{\c{s}}}, I.~C. and {Marka}, S. and {Marka}, Z. and {Maruyama}, R. and {Mase}, K. and {Maunu}, R. and {McNally}, F. and {Meagher}, K. and {Medina}, A. and {Meier}, M. and {Meighen-Berger}, S. and {Merz}, J. and {Meyers}, Z.~S. and {Micallef}, J. and {Mockler}, D. and {Moment{\'e}}, G. and {Montaruli}, T. and {Moore}, R.~W. and {Morse}, R. and {Moulai}, M. and {Muth}, P. and {Naab}, R. and {Nagai}, R. and {Nam}, J. and {Nauman}, U. and {Necker}, J. and {Neer}, G. and {Nelles}, A. and {Nguyễn}, L.~V. and {Niederhausen}, H. and {Nisa}, M.~U. and {Nowicki}, S.~C. and {Nygren}, D.~R. and {Oberla}, E. and {Pollmann}, A. Obertacke and {Oehler}, M. and {Olivas}, A. and {O'Sullivan}, E. and {Pan}, Y. and {Pandya}, H. and {Pankova}, D.~V. and {Papp}, L. and {Park}, N. and {Parker}, G.~K. and {Paudel}, E.~N. and {Peiffer}, P. and {P{\'e}rez de los Heros}, C. and {Petersen}, T.~C. and {Philippen}, S. and {Pieloth}, D. and {Pieper}, S. and {Pinfold}, J.~L. and {Pizzuto}, A. and {Plaisier}, I. and {Plum}, M. and {Popovych}, Y. and {Porcelli}, A. and {Rodriguez}, M. Prado and {Price}, P.~B. and {Przybylski}, G.~T. and {Raab}, C. and {Raissi}, A. and {Rameez}, M. and {Rauch}, L. and {Rawlins}, K. and {Rea}, I.~C. and {Rehman}, A. and {Reimann}, R. and {Renschler}, M. and {Renzi}, G. and {Resconi}, E. and {Reusch}, S. and {Rhode}, W. and {Richman}, M. and {Riedel}, B. and {Riegel}, M. and {Roberts}, E.~J. and {Robertson}, S. and {Roellinghoff}, G. and {Rongen}, M. and {Rott}, C. and {Ruhe}, T. and {Ryckbosch}, D. and {Cantu}, D. Rysewyk and {Safa}, I. and {Herrera}, S.~E. Sanchez and {Sandrock}, A. and {Sandroos}, J. and {Sandstrom}, P. and {Santander}, M. and {Sarkar}, S. and {Sarkar}, S. and {Satalecka}, K. and {Scharf}, M. and {Schaufel}, M. and {Schieler}, H. and {Schlunder}, P. and {Schmidt}, T. and {Schneider}, A. and {Schneider}, J. and {Schr{\"o}der}, F.~G. and {Schumacher}, L. and {Sclafani}, S. and {Seckel}, D. and {Seunarine}, S. and {Shaevitz}, M.~H. and {Sharma}, A. and {Shefali}, S. and {Silva}, M. and {Smith}, D. and {Smithers}, B. and {Snihur}, R. and {Soedingrekso}, J. and {Soldin}, D. and {S{\"o}ldner-Rembold}, S. and {Song}, M. and {Southall}, D. and {Spiczak}, G.~M. and {Spiering}, C. and {Stachurska}, J. and {Stamatikos}, M. and {Stanev}, T. and {Stein}, R. and {Stettner}, J. and {Steuer}, A. and {Stezelberger}, T. and {Stokstad}, R.~G. and {Strotjohann}, N.~L. and {St{\"u}rwald}, T. and {Stuttard}, T. and {Sullivan}, G.~W. and {Taboada}, I. and {Taketa}, A. and {Tanaka}, H.~K.~M. and {Tenholt}, F. and {Ter-Antonyan}, S. and {Terliuk}, A. and {Tilav}, S. and {Tollefson}, K. and {Tomankova}, L. and {T{\"o}nnis}, C. and {Torres}, J. and {Toscano}, S. and {Tosi}, D. and {Trettin}, A. and {Tselengidou}, M. and {Tung}, C.~F. and {Turcati}, A. and {Turcotte}, R. and {Turley}, C.~F. and {Twagirayezu}, J.~P. and {Ty}, B. and {Unger}, E. and {Elorrieta}, M.~A. Unland and {Vandenbroucke}, J. and {van Eijk}, D. and {van Eijndhoven}, N. and {Vannerom}, D. and {van Santen}, J. and {Veberic}, D. and {Verpoest}, S. and {Vieregg}, A. and {Vraeghe}, M. and {Walck}, C. and {Watson}, T.~B. and {Weaver}, C. and {Weindl}, A. and {Weinstock}, L. and {Weiss}, M.~J. and {Weldert}, J. and {Welling}, C. and {Wendt}, C. and {Werthebach}, J. and {Whitehorn}, N. and {Wiebe}, K. and {Wiebusch}, C.~H. and {Williams}, D.~R. and {Wissel}, S.~A. and {Wolf}, M. and {Wood}, T.~R. and {Woschnagg}, K. and {Wrede}, G. and {Wren}, S. and {Wulff}, J. and {Xu}, X.~W. and {Xu}, Y. and {Yanez}, J.~P. and {Yoshida}, S. and {Yuan}, T. and {Zhang}, Z. and {Zierke}, S. and {Z{\"o}cklein}, M.},
        title = "{IceCube-Gen2: the window to the extreme Universe}",
      journal = {Journal of Physics G Nuclear Physics},
         year = 2021,
        month = jun,
       volume = {48},
       number = {6},
          eid = {060501},
        pages = {060501},
          doi = {10.1088/1361-6471/abbd48},
archivePrefix = {arXiv},
       eprint = {2008.04323},
 primaryClass = {astro-ph.HE},
       adsurl = {https://ui.adsabs.harvard.edu/abs/2021JPhG...48f0501A}
}

@ARTICLE{2022ApJ...931...59P,
       author = {{Peirson}, Abel L. and {Liodakis}, Ioannis and {Romani}, Roger W.},
        title = "{Testing High-energy Emission Models for Blazars with X-Ray Polarimetry}",
      journal = {\apj},
         year = 2022,
        month = may,
       volume = {931},
       number = {1},
          eid = {59},
        pages = {59},
          doi = {10.3847/1538-4357/ac6a54},
archivePrefix = {arXiv},
       eprint = {2204.11803},
 primaryClass = {astro-ph.HE},
       adsurl = {https://ui.adsabs.harvard.edu/abs/2022ApJ...931...59P}
}

@ARTICLE{Bloom1996,
   author = {{Bloom}, S.~D. and {Marscher}, A.~P.},
    title = "{An Analysis of the Synchrotron Self-Compton Model for the Multi--Wave Band Spectra of Blazars}",
  journal = {\apj},
     year = 1996,
    month = apr,
   volume = 461,
    pages = {657},
      doi = {10.1086/177092},
   adsurl = {http://adsabs.harvard.edu/abs/1996ApJ...461..657B}
}

@article{Sokolov_2004,
doi = {10.1086/423165},
url = {https://dx.doi.org/10.1086/423165},
year = {2004},
month = {oct},
publisher = {},
volume = {613},
number = {2},
pages = {725},
author = {Andrei Sokolov and Alan P. Marscher and Ian M. McHardy},
title = {Synchrotron Self-Compton Model for Rapid Nonthermal Flares in Blazars with Frequency-dependent Time Lags},
journal = {The Astrophysical Journal}
}

@ARTICLE{1992A&A...256L..27D,
       author = {{Dermer}, C.~D. and {Schlickeiser}, R. and {Mastichiadis}, A.},
        title = "{High-energy gamma radiation from extragalactic radio sources.}",
      journal = {\aap},
         year = 1992,
        month = mar,
       volume = {256},
        pages = {L27-L30},
       adsurl = {https://ui.adsabs.harvard.edu/abs/1992A&A...256L..27D}
}

@ARTICLE{1993ApJ...416..458D,
       author = {{Dermer}, Charles D. and {Schlickeiser}, Reinhard},
        title = "{Model for the High-Energy Emission from Blazars}",
      journal = {\apj},
         year = 1993,
        month = oct,
       volume = {416},
        pages = {458},
          doi = {10.1086/173251},
       adsurl = {https://ui.adsabs.harvard.edu/abs/1993ApJ...416..458D}
}

@ARTICLE{1994ApJS...90..923S,
       author = {{Sikora}, Marek},
        title = "{High-Energy Radiation from Active Galactic Nuclei}",
      journal = {\apjs},
         year = 1994,
        month = feb,
       volume = {90},
        pages = {923},
          doi = {10.1086/191926},
       adsurl = {https://ui.adsabs.harvard.edu/abs/1994ApJS...90..923S}
}

@ARTICLE{2000ApJ...545..107B,
       author = {{B{\l}a{\.z}ejowski}, M. and {Sikora}, M. and {Moderski}, R. and {Madejski}, G.~M.},
        title = "{Comptonization of Infrared Radiation from Hot Dust by Relativistic Jets in Quasars}",
      journal = {\apj},
         year = 2000,
        month = dec,
       volume = {545},
       number = {1},
        pages = {107-116},
          doi = {10.1086/317791},
archivePrefix = {arXiv},
       eprint = {astro-ph/0008154},
 primaryClass = {astro-ph},
       adsurl = {https://ui.adsabs.harvard.edu/abs/2000ApJ...545..107B}
}

@ARTICLE{2009ApJ...704...38S,
       author = {{Sikora}, Marek and {Stawarz}, {\L}ukasz and {Moderski}, Rafa{\l} and {Nalewajko}, Krzysztof and {Madejski}, Greg M.},
        title = "{Constraining Emission Models of Luminous Blazar Sources}",
      journal = {\apj},
         year = 2009,
        month = oct,
       volume = {704},
       number = {1},
        pages = {38-50},
          doi = {10.1088/0004-637X/704/1/38},
archivePrefix = {arXiv},
       eprint = {0904.1414},
 primaryClass = {astro-ph.CO},
       adsurl = {https://ui.adsabs.harvard.edu/abs/2009ApJ...704...38S}
}

@article{Agudo_2012,
doi = {10.1088/1742-6596/355/1/012032},
url = {https://dx.doi.org/10.1088/1742-6596/355/1/012032},
year = {2012},
month = {mar},
publisher = {},
volume = {355},
number = {1},
pages = {012032},
author = {I Agudo and S G Jorstad and A P Marscher and V M Larionov and J L Gómez and A Lähteenmäki and M Gurwell and P S Smith and H Wiesemeyer and C Thum and J Heidt},
title = {γ-ray emission region located in the parsec scale jet of OJ287},
journal = {Journal of Physics: Conference Series}
}

@ARTICLE{1999PASA...16..160M,
       author = {{M{\"u}cke}, A. and {Rachen}, J.~P. and {Engel}, Ralph and {Protheroe}, R.~J. and {Stanev}, Todor},
        title = "{Photohadronic Processes in Astrophysical Environments}",
      journal = {\pasa},
         year = 1999,
        month = aug,
       volume = {16},
       number = {2},
        pages = {160-166},
          doi = {10.1071/AS99160},
archivePrefix = {arXiv},
       eprint = {astro-ph/9808279},
 primaryClass = {astro-ph},
       adsurl = {https://ui.adsabs.harvard.edu/abs/1999PASA...16..160M}
}

@ARTICLE{2003APh....18..593M,
       author = {{M{\"u}cke}, A. and {Protheroe}, R.~J. and {Engel}, R. and {Rachen}, J.~P. and {Stanev}, T.},
        title = "{BL Lac objects in the synchrotron proton blazar model}",
      journal = {Astroparticle Physics},
         year = 2003,
        month = mar,
       volume = {18},
       number = {6},
        pages = {593-613},
          doi = {10.1016/S0927-6505(02)00185-8},
archivePrefix = {arXiv},
       eprint = {astro-ph/0206164},
 primaryClass = {astro-ph},
       adsurl = {https://ui.adsabs.harvard.edu/abs/2003APh....18..593M}
}

@INPROCEEDINGS{2001AIPC..558..700P,
       author = {{Protheroe}, Raymond J. and {M{\"u}cke}, Anita},
        title = "{Application of the synchrotron proton blazar model to BL Lac objects}",
    booktitle = {High Energy Gamma-Ray Astronomy: International Symposium},
         year = 2001,
       editor = {{Aharonian}, Felix A. and {V{\"o}lk}, Heinz J.},
       series = {American Institute of Physics Conference Series},
       volume = {558},
        month = apr,
        pages = {700-703},
          doi = {10.1063/1.1370856},
archivePrefix = {arXiv},
       eprint = {astro-ph/0011154},
 primaryClass = {astro-ph},
       adsurl = {https://ui.adsabs.harvard.edu/abs/2001AIPC..558..700P}
}

@ARTICLE{2005ApJ...630..186R,
       author = {{Reimer}, A. and {B{\"o}ttcher}, M. and {Postnikov}, S.},
        title = "{Neutrino Emission in the Hadronic Synchrotron Mirror Model: The ``Orphan'' TeV Flare from 1ES 1959+650}",
      journal = {\apj},
         year = 2005,
        month = sep,
       volume = {630},
       number = {1},
        pages = {186-190},
          doi = {10.1086/431948},
archivePrefix = {arXiv},
       eprint = {astro-ph/0505233},
 primaryClass = {astro-ph},
       adsurl = {https://ui.adsabs.harvard.edu/abs/2005ApJ...630..186R}
}

@ARTICLE{2020Galax...8...72C,
       author = {{Cerruti}, Matteo},
        title = "{Leptonic and Hadronic Radiative Processes in Supermassive-Black-Hole Jets}",
      journal = {Galaxies},
         year = 2020,
        month = oct,
       volume = {8},
       number = {4},
        pages = {72},
          doi = {10.3390/galaxies8040072},
archivePrefix = {arXiv},
       eprint = {2012.13302},
 primaryClass = {astro-ph.HE},
       adsurl = {https://ui.adsabs.harvard.edu/abs/2020Galax...8...72C}
}

@ARTICLE{2000NuPhS..80C0810M,
       author = {{M{\"u}cke}, A. and {Rachen}, J.~P. and {Engel}, R. and {Protheroe}, R.~J. and {Stanev}, T.},
        title = "{Photomeson production in astrophysical sources}",
      journal = {Nuclear Physics B Proceedings Supplements},
         year = 2000,
        month = jan,
       volume = {80},
        pages = {08/10},
          doi = {10.48550/arXiv.astro-ph/9905153},
archivePrefix = {arXiv},
       eprint = {astro-ph/9905153},
 primaryClass = {astro-ph},
       adsurl = {https://ui.adsabs.harvard.edu/abs/2000NuPhS..80C0810M}
}

@ARTICLE{2018Sci...361.1378I,
       author = {{IceCube Collaboration} and {Aartsen}, M.~G. and {Ackermann}, M. and {Adams}, J. and {Aguilar}, J.~A. and {Ahlers}, M. and {Ahrens}, M. and {Al Samarai}, I. and {Altmann}, D. and {Andeen}, K. and {Anderson}, T. and {Ansseau}, I. and {Anton}, G. and {Arg{\"u}elles}, C. and {Auffenberg}, J. and {Axani}, S. and {Bagherpour}, H. and {Bai}, X. and {Barron}, J.~P. and {Barwick}, S.~W. and {Baum}, V. and {Bay}, R. and {Beatty}, J.~J. and {Becker Tjus}, J. and {Becker}, K. -H. and {BenZvi}, S. and {Berley}, D. and {Bernardini}, E. and {Besson}, D.~Z. and {Binder}, G. and {Bindig}, D. and {Blaufuss}, E. and {Blot}, S. and {Bohm}, C. and {B{\"o}rner}, M. and {Bos}, F. and {B{\"o}ser}, S. and {Botner}, O. and {Bourbeau}, E. and {Bourbeau}, J. and {Bradascio}, F. and {Braun}, J. and {Brenzke}, M. and {Bretz}, H. -P. and {Bron}, S. and {Brostean-Kaiser}, J. and {Burgman}, A. and {Busse}, R.~S. and {Carver}, T. and {Cheung}, E. and {Chirkin}, D. and {Christov}, A. and {Clark}, K. and {Classen}, L. and {Coenders}, S. and {Collin}, G.~H. and {Conrad}, J.~M. and {Coppin}, P. and {Correa}, P. and {Cowen}, D.~F. and {Cross}, R. and {Dave}, P. and {Day}, M. and {de Andr{\'e}}, J.~P.~A.~M. and {De Clercq}, C. and {DeLaunay}, J.~J. and {Dembinski}, H. and {De Ridder}, S. and {Desiati}, P. and {de Vries}, K.~D. and {de Wasseige}, G. and {de With}, M. and {DeYoung}, T. and {D{\'\i}az-V{\'e}lez}, J.~C. and {di Lorenzo}, V. and {Dujmovic}, H. and {Dumm}, J.~P. and {Dunkman}, M. and {Dvorak}, E. and {Eberhardt}, B. and {Ehrhardt}, T. and {Eichmann}, B. and {Eller}, P. and {Evenson}, P.~A. and {Fahey}, S. and {Fazely}, A.~R. and {Felde}, J. and {Filimonov}, K. and {Finley}, C. and {Flis}, S. and {Franckowiak}, A. and {Friedman}, E. and {Fritz}, A. and {Gaisser}, T.~K. and {Gallagher}, J. and {Gerhardt}, L. and {Ghorbani}, K. and {Glauch}, T. and {Gl{\"u}senkamp}, T. and {Goldschmidt}, A. and {Gonzalez}, J.~G. and {Grant}, D. and {Griffith}, Z. and {Haack}, C. and {Hallgren}, A. and {Halzen}, F. and {Hanson}, K. and {Hebecker}, D. and {Heereman}, D. and {Helbing}, K. and {Hellauer}, R. and {Hickford}, S. and {Hignight}, J. and {Hill}, G.~C. and {Hoffman}, K.~D. and {Hoffmann}, R. and {Hoinka}, T. and {Hokanson-Fasig}, B. and {Hoshina}, K. and {Huang}, F. and {Huber}, M. and {Hultqvist}, K. and {H{\"u}nnefeld}, M. and {Hussain}, R. and {In}, S. and {Iovine}, N. and {Ishihara}, A. and {Jacobi}, E. and {Japaridze}, G.~S. and {Jeong}, M. and {Jero}, K. and {Jones}, B.~J.~P. and {Kalaczynski}, P. and {Kang}, W. and {Kappes}, A. and {Kappesser}, D. and {Karg}, T. and {Karle}, A. and {Katz}, U. and {Kauer}, M. and {Keivani}, A. and {Kelley}, J.~L. and {Kheirandish}, A. and {Kim}, J. and {Kim}, M. and {Kintscher}, T. and {Kiryluk}, J. and {Kittler}, T. and {Klein}, S.~R. and {Koirala}, R. and {Kolanoski}, H. and {K{\"o}pke}, L. and {Kopper}, C. and {Kopper}, S. and {Koschinsky}, J.~P. and {Koskinen}, D.~J. and {Kowalski}, M. and {Krings}, K. and {Kroll}, M. and {Kr{\"u}ckl}, G. and {Kunwar}, S. and {Kurahashi}, N. and {Kuwabara}, T. and {Kyriacou}, A. and {Labare}, M. and {Lanfranchi}, J.~L. and {Larson}, M.~J. and {Lauber}, F. and {Leonard}, K. and {Lesiak-Bzdak}, M. and {Leuermann}, M. and {Liu}, Q.~R. and {Lozano Mariscal}, C.~J. and {Lu}, L. and {L{\"u}nemann}, J. and {Luszczak}, W. and {Madsen}, J. and {Maggi}, G. and {Mahn}, K.~B.~M. and {Mancina}, S. and {Maruyama}, R. and {Mase}, K. and {Maunu}, R. and {Meagher}, K. and {Medici}, M. and {Meier}, M. and {Menne}, T. and {Merino}, G. and {Meures}, T. and {Miarecki}, S. and {Micallef}, J. and {Moment{\'e}}, G. and {Montaruli}, T. and {Moore}, R.~W. and {Morse}, R. and {Moulai}, M. and {Nahnhauer}, R. and {Nakarmi}, P. and {Naumann}, U. and {Neer}, G. and {Niederhausen}, H. and {Nowicki}, S.~C. and {Nygren}, D.~R. and {Obertacke Pollmann}, A. and {Olivas}, A. and {O'Murchadha}, A. and {O'Sullivan}, E. and {Palczewski}, T. and {Pandya}, H. and {Pankova}, D.~V. and {Peiffer}, P. and {Pepper}, J.~A. and {P{\'e}rez de los Heros}, C. and {Pieloth}, D. and {Pinat}, E. and {Plum}, M. and {Price}, P.~B. and {Przybylski}, G.~T. and {Raab}, C. and {R{\"a}del}, L. and {Rameez}, M. and {Rauch}, L. and {Rawlins}, K. and {Rea}, I.~C. and {Reimann}, R. and {Relethford}, B. and {Relich}, M. and {Resconi}, E. and {Rhode}, W. and {Richman}, M. and {Robertson}, S. and {Rongen}, M. and {Rott}, C. and {Ruhe}, T. and {Ryckbosch}, D. and {Rysewyk}, D. and {Safa}, I. and {S{\"a}lzer}, T. and {Sanchez Herrera}, S.~E. and {Sandrock}, A. and {Sandroos}, J. and {Santander}, M. and {Sarkar}, S. and {Sarkar}, S. and {Satalecka}, K. and {Schlunder}, P. and {Schmidt}, T. and {Schneider}, A. and {Schoenen}, S. and {Sch{\"o}neberg}, S. and {Schumacher}, L. and {Sclafani}, S. and {Seckel}, D. and {Seunarine}, S. and {Soedingrekso}, J. and {Soldin}, D. and {Song}, M. and {Spiczak}, G.~M. and {Spiering}, C. and {Stachurska}, J. and {Stamatikos}, M. and {Stanev}, T. and {Stasik}, A. and {Stein}, R. and {Stettner}, J. and {Steuer}, A. and {Stezelberger}, T. and {Stokstad}, R.~G. and {St{\"o}{\ss}l}, A. and {Strotjohann}, N.~L. and {Stuttard}, T. and {Sullivan}, G.~W. and {Sutherland}, M. and {Taboada}, I. and {Tatar}, J. and {Tenholt}, F. and {Ter-Antonyan}, S. and {Terliuk}, A. and {Tilav}, S. and {Toale}, P.~A. and {Tobin}, M.~N. and {Toennis}, C. and {Toscano}, S. and {Tosi}, D. and {Tselengidou}, M. and {Tung}, C.~F. and {Turcati}, A. and {Turley}, C.~F. and {Ty}, B. and {Unger}, E. and {Usner}, M. and {Vandenbroucke}, J. and {Van Driessche}, W. and {van Eijk}, D. and {van Eijndhoven}, N. and {Vanheule}, S. and {van Santen}, J. and {Vogel}, E. and {Vraeghe}, M. and {Walck}, C. and {Wallace}, A. and {Wallraff}, M. and {Wandler}, F.~D. and {Wandkowsky}, N. and {Waza}, A. and {Weaver}, C. and {Weiss}, M.~J. and {Wendt}, C. and {Werthebach}, J. and {Westerhoff}, S. and {Whelan}, B.~J. and {Whitehorn}, N. and {Wiebe}, K. and {Wiebusch}, C.~H. and {Wille}, L. and {Williams}, D.~R. and {Wills}, L. and {Wolf}, M. and {Wood}, J. and {Wood}, T.~R. and {Woschnagg}, K. and {Xu}, D.~L. and {Xu}, X.~W. and {Xu}, Y. and {Yanez}, J.~P. and {Yodh}, G. and {Yoshida}, S. and {Yuan}, T. and {Fermi-LAT Collaboration} and {Abdollahi}, S. and {Ajello}, M. and {Angioni}, R. and {Baldini}, L. and {Ballet}, J. and {Barbiellini}, G. and {Bastieri}, D. and {Bechtol}, K. and {Bellazzini}, R. and {Berenji}, B. and {Bissaldi}, E. and {Blandford}, R.~D. and {Bonino}, R. and {Bottacini}, E. and {Bregeon}, J. and {Bruel}, P. and {Buehler}, R. and {Burnett}, T.~H. and {Burns}, E. and {Buson}, S. and {Cameron}, R.~A. and {Caputo}, R. and {Caraveo}, P.~A. and {Cavazzuti}, E. and {Charles}, E. and {Chen}, S. and {Cheung}, C.~C. and {Chiang}, J. and {Chiaro}, G. and {Ciprini}, S. and {Cohen-Tanugi}, J. and {Conrad}, J. and {Costantin}, D. and {Cutini}, S. and {D'Ammando}, F. and {de Palma}, F. and {Digel}, S.~W. and {Di Lalla}, N. and {Di Mauro}, M. and {Di Venere}, L. and {Dom{\'\i}nguez}, A. and {Favuzzi}, C. and {Franckowiak}, A. and {Fukazawa}, Y. and {Funk}, S. and {Fusco}, P. and {Gargano}, F. and {Gasparrini}, D. and {Giglietto}, N. and {Giomi}, M. and {Giommi}, P. and {Giordano}, F. and {Giroletti}, M. and {Glanzman}, T. and {Green}, D. and {Grenier}, I.~A. and {Grondin}, M. -H. and {Guiriec}, S. and {Harding}, A.~K. and {Hayashida}, M. and {Hays}, E. and {Hewitt}, J.~W. and {Horan}, D. and {J{\'o}hannesson}, G. and {Kadler}, M. and {Kensei}, S. and {Kocevski}, D. and {Krauss}, F. and {Kreter}, M. and {Kuss}, M. and {La Mura}, G. and {Larsson}, S. and {Latronico}, L. and {Lemoine-Goumard}, M. and {Li}, J. and {Longo}, F. and {Loparco}, F. and {Lovellette}, M.~N. and {Lubrano}, P. and {Magill}, J.~D. and {Maldera}, S. and {Malyshev}, D. and {Manfreda}, A. and {Mazziotta}, M.~N. and {McEnery}, J.~E. and {Meyer}, M. and {Michelson}, P.~F. and {Mizuno}, T. and {Monzani}, M.~E. and {Morselli}, A. and {Moskalenko}, I.~V. and {Negro}, M. and {Nuss}, E. and {Ojha}, R. and {Omodei}, N. and {Orienti}, M. and {Orlando}, E. and {Palatiello}, M. and {Paliya}, V.~S. and {Perkins}, J.~S. and {Persic}, M. and {Pesce-Rollins}, M. and {Piron}, F. and {Porter}, T.~A. and {Principe}, G. and {Rain{\`o}}, S. and {Rando}, R. and {Rani}, B. and {Razzano}, M. and {Razzaque}, S. and {Reimer}, A. and {Reimer}, O. and {Renault-Tinacci}, N. and {Ritz}, S. and {Rochester}, L.~S. and {Saz Parkinson}, P.~M. and {Sgr{\`o}}, C. and {Siskind}, E.~J. and {Spandre}, G. and {Spinelli}, P. and {Suson}, D.~J. and {Tajima}, H. and {Takahashi}, M. and {Tanaka}, Y. and {Thayer}, J.~B. and {Thompson}, D.~J. and {Tibaldo}, L. and {Torres}, D.~F. and {Torresi}, E. and {Tosti}, G. and {Troja}, E. and {Valverde}, J. and {Vianello}, G. and {Vogel}, M. and {Wood}, K. and {Wood}, M. and {Zaharijas}, G. and {MAGIC Collaboration} and {Ahnen}, M.~L. and {Ansoldi}, S. and {Antonelli}, L.~A. and {Arcaro}, C. and {Baack}, D. and {Babi{\'c}}, A. and {Banerjee}, B. and {Bangale}, P. and {Barres de Almeida}, U. and {Barrio}, J.~A. and {Becerra Gonz{\'a}lez}, J. and {Bednarek}, W. and {Bernardini}, E. and {Berti}, A. and {Bhattacharyya}, W. and {Biland}, A. and {Blanch}, O. and {Bonnoli}, G. and {Carosi}, A. and {Carosi}, R. and {Ceribella}, G. and {Chatterjee}, A. and {Colak}, S.~M. and {Colin}, P. and {Colombo}, E. and {Contreras}, J.~L. and {Cortina}, J. and {Covino}, S. and {Cumani}, P. and {Da Vela}, P. and {Dazzi}, F. and {De Angelis}, A. and {De Lotto}, B. and {Delfino}, M. and {Delgado}, J. and {Di Pierro}, F. and {Dom{\'\i}nguez}, A. and {Dominis Prester}, D. and {Dorner}, D. and {Doro}, M. and {Einecke}, S. and {Elsaesser}, D. and {Fallah Ramazani}, V. and {Fern{\'a}ndez-Barral}, A. and {Fidalgo}, D. and {Foffano}, L. and {Pfrang}, K. and {Fonseca}, M.~V. and {Font}, L. and {Franceschini}, A. and {Fruck}, C. and {Galindo}, D. and {Gallozzi}, S. and {Garc{\'\i}a L{\'o}pez}, R.~J. and {Garczarczyk}, M. and {Gaug}, M. and {Giammaria}, P. and {Godinovi{\'c}}, N. and {Gora}, D. and {Guberman}, D. and {Hadasch}, D. and {Hahn}, A. and {Hassan}, T. and {Hayashida}, M. and {Herrera}, J. and {Hose}, J. and {Hrupec}, D. and {Inoue}, S. and {Ishio}, K. and {Konno}, Y. and {Kubo}, H. and {Kushida}, J. and {Lelas}, D. and {Lindfors}, E. and {Lombardi}, S. and {Longo}, F. and {L{\'o}pez}, M. and {Maggio}, C. and {Majumdar}, P. and {Makariev}, M. and {Maneva}, G. and {Manganaro}, M. and {Mannheim}, K. and {Maraschi}, L. and {Mariotti}, M. and {Mart{\'\i}nez}, M. and {Masuda}, S. and {Mazin}, D. and {Minev}, M. and {M}, J.~M. and {Mirzoyan}, R. and {Moralejo}, A. and {Moreno}, V. and {Moretti}, E. and {Nagayoshi}, T. and {Neustroev}, V. and {Niedzwiecki}, A. and {Nievas Rosillo}, M. and {Nigro}, C. and {Nilsson}, K. and {Ninci}, D. and {Nishijima}, K. and {Noda}, K. and {Nogu{\'e}s}, L. and {Paiano}, S. and {Palacio}, J. and {Paneque}, D. and {Paoletti}, R. and {Paredes}, J.~M. and {Pedaletti}, G. and {Peresano}, M. and {Persic}, M. and {Prada Moroni}, P.~G. and {Prandini}, E. and {Puljak}, I. and {Rodriguez Garcia}, J. and {Reichardt}, I. and {Rhode}, W. and {Rib{\'o}}, M. and {Rico}, J. and {Righi}, C. and {Rugliancich}, A. and {Saito}, T. and {Satalecka}, K. and {Schweizer}, T. and {Sitarek}, J. and {{\v{S}}nidaric {\textasciiacute}}, I. and {Sobczynska}, D. and {Stamerra}, A. and {Strzys}, M. and {Suri{\'c}}, T. and {Takahashi}, M. and {Tavecchio}, F. and {Temnikov}, P. and {Terzi{\'c}}, T. and {Teshima}, M. and {Torres-Alb{\`a}}, N. and {Treves}, A. and {Tsujimoto}, S. and {Vanzo}, G. and {Vazquez Acosta}, M. and {Vovk}, I. and {Ward}, J.~E. and {Will}, M. and {S} and {Zaric {\textasciiacute}}, D. and {AGILE Team} and {Lucarelli}, F. and {Tavani}, M. and {Piano}, G. and {Donnarumma}, I. and {Pittori}, C. and {Verrecchia}, F. and {Barbiellini}, G. and {Bulgarelli}, A. and {Caraveo}, P. and {Cattaneo}, P.~W. and {Colafrancesco}, S. and {Costa}, E. and {Di Cocco}, G. and {Ferrari}, A. and {Gianotti}, F. and {Giuliani}, A. and {Lipari}, P. and {Mereghetti}, S. and {Morselli}, A. and {Pacciani}, L. and {Paoletti}, F. and {Parmiggiani}, N. and {Pellizzoni}, A. and {Picozza}, P. and {Pilia}, M. and {Rappoldi}, A. and {Trois}, A. and {Vercellone}, S. and {Vittorini}, V. and {ASAS-SN Team} and {Stanek}, K.~Z. and {Franckowiak}, A. and {Kochanek}, C.~S. and {Beacom}, J.~F. and {Thompson}, T.~A. and {Holoien}, T.~W. -S. and {Dong}, S. and {Prieto}, J.~L. and {Shappee}, B.~J. and {Holmbo}, S. and {HAWC Collaboration} and {Abeysekara}, A.~U. and {Albert}, A. and {Alfaro}, R. and {Alvarez}, C. and {Arceo}, R. and {Arteaga-Vel{\'a}zquez}, J.~C. and {Avila Rojas}, D. and {Ayala Solares}, H.~A. and {Becerril}, A. and {Belmont-Moreno}, E. and {Bernal}, A. and {Caballero-Mora}, K.~S. and {Capistr{\'a}n}, T. and {Carrami{\~n}ana}, A. and {Casanova}, S. and {Castillo}, M. and {Cotti}, U. and {Cotzomi}, J. and {Couti{\~n}o de Le{\'o}n}, S. and {De Le{\'o}n}, C. and {De la Fuente}, E. and {Diaz Hernandez}, R. and {Dichiara}, S. and {Dingus}, B.~L. and {DuVernois}, M.~A. and {D{\'\i}az-V{\'e}lez}, J.~C. and {Ellsworth}, R.~W. and {Engel}, K. and {Fiorino}, D.~W. and {Fleischhack}, H. and {Fraija}, N. and {Garc{\'\i}a-Gonz{\'a}lez}, J.~A. and {Garfias}, F. and {Gonz{\'a}lez Mu{\~n}oz}, A. and {Gonz{\'a}lez}, M.~M. and {Goodman}, J.~A. and {Hampel-Arias}, Z. and {Harding}, J.~P. and {Hernandez}, S. and {Hona}, B. and {Hueyotl-Zahuantitla}, F. and {Hui}, C.~M. and {H{\"u}ntemeyer}, P. and {Iriarte}, A. and {Jardin-Blicq}, A. and {Joshi}, V. and {Kaufmann}, S. and {Kunde}, G.~J. and {Lara}, A. and {Lauer}, R.~J. and {Lee}, W.~H. and {Lennarz}, D. and {Le{\'o}n Vargas}, H. and {Linnemann}, J.~T. and {Longinotti}, A.~L. and {Luis-Raya}, G. and {Luna-Garc{\'\i}a}, R. and {Malone}, K. and {Marinelli}, S.~S. and {Martinez}, O. and {Martinez-Castellanos}, I. and {Mart{\'\i}nez-Castro}, J. and {Mart{\'\i}nez-Huerta}, H. and {Matthews}, J.~A. and {Miranda-Romagnoli}, P. and {Moreno}, E. and {Mostaf{\'a}}, M. and {Nayerhoda}, A. and {Nellen}, L. and {Newbold}, M. and {Nisa}, M.~U. and {Noriega-Papaqui}, R. and {Pelayo}, R. and {Pretz}, J. and {P{\'e}rez-P{\'e}rez}, E.~G. and {Ren}, Z. and {Rho}, C.~D. and {Rivi{\`e}re}, C. and {Rosa-Gonz{\'a}lez}, D. and {Rosenberg}, M. and {Ruiz-Velasco}, E. and {Ruiz-Velasco}, E. and {Salesa Greus}, F. and {Sandoval}, A. and {Schneider}, M. and {Schoorlemmer}, H. and {Sinnis}, G. and {Smith}, A.~J. and {Springer}, R.~W. and {Surajbali}, P. and {Tibolla}, O. and {Tollefson}, K. and {Torres}, I. and {Villase{\~n}or}, L. and {Weisgarber}, T. and {Werner}, F. and {Yapici}, T. and {Gaurang}, Y. and {Zepeda}, A. and {Zhou}, H. and {{\'A}lvarez}, J.~D. and {H.~E.~S.~S. Collaboration} and {Abdalla}, H. and {Ang{\"u}ner}, E.~O. and {Armand}, C. and {Backes}, M. and {Becherini}, Y. and {Berge}, D. and {B{\"o}ttcher}, M. and {Boisson}, C. and {Bolmont}, J. and {Bonnefoy}, S. and {Bordas}, P. and {Brun}, F. and {B{\"u}chele}, M. and {Bulik}, T. and {Caroff}, S. and {Carosi}, A. and {Casanova}, S. and {Cerruti}, M. and {Chakraborty}, N. and {Chandra}, S. and {Chen}, A. and {Colafrancesco}, S. and {Davids}, I.~D. and {Deil}, C. and {Devin}, J. and {Djannati-Ata{\"\i}}, A. and {Egberts}, K. and {Emery}, G. and {Eschbach}, S. and {Fiasson}, A. and {Fontaine}, G. and {Funk}, S. and {F{\"u}{\ss}ling}, M. and {Gallant}, Y.~A. and {Gat{\'e}}, F. and {Giavitto}, G. and {Glawion}, D. and {Glicenstein}, J.~F. and {Gottschall}, D. and {Grondin}, M. -H. and {Haupt}, M. and {Henri}, G. and {Hinton}, J.~A. and {Hoischen}, C. and {Holch}, T.~L. and {Huber}, D. and {Jamrozy}, M. and {Jankowsky}, D. and {Jankowsky}, F. and {Jouvin}, L. and {Jung-Richardt}, I. and {Kerszberg}, D. and {Kh{\'e}lifi}, B. and {King}, J. and {Klepser}, S. and {Kluz {\textasciiacute}niak}, W. and {Komin}, Nu. and {Kraus}, M. and {Lefaucheur}, J. and {Lemi{\`e}re}, A. and {Lemoine-Goumard}, M. and {Lenain}, J. -P. and {Leser}, E. and {Lohse}, T. and {L{\'o}pez-Coto}, R. and {Lorentz}, M. and {Lypova}, I. and {Marandon}, V. and {Guillem Mart{\'\i}-Devesa}, G. and {Maurin}, G. and {Mitchell}, A.~M.~W. and {Moderski}, R. and {Mohamed}, M. and {Mohrmann}, L. and {Moulin}, E. and {Murach}, T. and {de Naurois}, M. and {Niederwanger}, F. and {Niemiec}, J. and {Oakes}, L. and {O'Brien}, P. and {Ohm}, S. and {Ostrowski}, M. and {Oya}, I. and {Panter}, M. and {Parsons}, R.~D. and {Perennes}, C. and {Piel}, Q. and {Pita}, S. and {Poireau}, V. and {Priyana Noel}, A. and {Prokoph}, H. and {P{\"u}hlhofer}, G. and {Quirrenbach}, A. and {Raab}, S. and {Rauth}, R. and {Renaud}, M. and {Rieger}, F. and {Rinchiuso}, L. and {Romoli}, C. and {Rowell}, G. and {Rudak}, B. and {Sasaki}, D.~A. and {Sanchez}, M. and {Schlickeiser}, R. and {Sch{\"u}ssler}, F. and {Schulz}, A. and {Schwanke}, U. and {Seglar-Arroyo}, M. and {Shafi}, N. and {Simoni}, R. and {Sol}, H. and {Stegmann}, C. and {Steppa}, C. and {Tavernier}, T. and {Taylor}, A.~M. and {Tiziani}, D. and {Trichard}, C. and {Tsirou}, M. and {van Eldik}, C. and {van Rensburg}, C. and {van Soelen}, B. and {Veh}, J. and {Vincent}, P. and {Voisin}, F. and {Wagner}, S.~J. and {Wagner}, R.~M. and {Wierzcholska}, A. and {Zanin}, R. and {Zdziarski}, A.~A. and {Zech}, A. and {Ziegler}, A. and {Zorn}, J. and {{\.Z}ywucka}, N. and {INTEGRAL Team} and {Savchenko}, V. and {Ferrigno}, C. and {Bazzano}, A. and {Diehl}, R. and {Kuulkers}, E. and {Laurent}, P. and {Mereghetti}, S. and {Natalucci}, L. and {Panessa}, F. and {Rodi}, J. and {Ubertini}, P. and {Kanata}, Kiso and Subaru Observing Teams and {Morokuma}, T. and {Ohta}, K. and {Tanaka}, Y.~T. and {Mori}, H. and {Yamanaka}, M. and {Kawabata}, K.~S. and {Utsumi}, Y. and {Nakaoka}, T. and {Kawabata}, M. and {Nagashima}, H. and {Yoshida}, M. and {Matsuoka}, Y. and {Itoh}, R. and {Kapteyn Team} and {Keel}, W. and {Liverpool Telescope Team} and {Copperwheat}, C. and {Steele}, I. and {Swift/NuSTAR Team} and {Cenko}, S.~B. and {Cowen}, D.~F. and {DeLaunay}, J.~J. and {Evans}, P.~A. and {Fox}, D.~B. and {Keivani}, A. and {Kennea}, J.~A. and {Marshall}, F.~E. and {Osborne}, J.~P. and {Santander}, M. and {Tohuvavohu}, A. and {Turley}, C.~F. and {VERITAS Collaboration} and {Abeysekara}, A.~U. and {Archer}, A. and {Benbow}, W. and {Bird}, R. and {Brill}, A. and {Brose}, R. and {Buchovecky}, M. and {Buckley}, J.~H. and {Bugaev}, V. and {Christiansen}, J.~L. and {Connolly}, M.~P. and {Cui}, W. and {Daniel}, M.~K. and {Errando}, M. and {Falcone}, A. and {Feng}, Q. and {Finley}, J.~P. and {Fortson}, L. and {Furniss}, A. and {Gueta}, O. and {H{\"u}tten}, M. and {Hervet}, O. and {Hughes}, G. and {Humensky}, T.~B. and {Johnson}, C.~A. and {Kaaret}, P. and {Kar}, P. and {Kelley-Hoskins}, N. and {Kertzman}, M. and {Kieda}, D. and {Krause}, M. and {Krennrich}, F. and {Kumar}, S. and {Lang}, M.~J. and {Lin}, T.~T.~Y. and {Maier}, G. and {McArthur}, S. and {Moriarty}, P. and {Mukherjee}, R. and {Nieto}, D. and {O'Brien}, S. and {Ong}, R.~A. and {Otte}, A.~N. and {Park}, N. and {Petrashyk}, A. and {Pohl}, M. and {Popkow}, A. and {Pueschel}, S.~E. and {Quinn}, J. and {Ragan}, K. and {Reynolds}, P.~T. and {Richards}, G.~T. and {Roache}, E. and {Rulten}, C. and {Sadeh}, I. and {Santander}, M. and {Scott}, S.~S. and {Sembroski}, G.~H. and {Shahinyan}, K. and {Sushch}, I. and {Tr{\'e}panier}, S. and {Tyler}, J. and {Vassiliev}, V.~V. and {Wakely}, S.~P. and {Weinstein}, A. and {Wells}, R.~M. and {Wilcox}, P. and {Wilhelm}, A. and {Williams}, D.~A. and {Zitzer}, B. and {VLA/B Team} and {Tetarenko}, A.~J. and {Kimball}, A.~E. and {Miller-Jones}, J.~C.~A. and {Sivakoff}, G.~R.},
        title = "{Multimessenger observations of a flaring blazar coincident with high-energy neutrino IceCube-170922A}",
      journal = {Science},
         year = 2018,
        month = jul,
       volume = {361},
       number = {6398},
          eid = {eaat1378},
        pages = {eaat1378},
          doi = {10.1126/science.aat1378},
archivePrefix = {arXiv},
       eprint = {1807.08816},
 primaryClass = {astro-ph.HE},
       adsurl = {https://ui.adsabs.harvard.edu/abs/2018Sci...361.1378I}
}

@ARTICLE{2024MNRAS.527L..26S,
       author = {{Suray}, Alisa and {Troitsky}, Sergey},
        title = "{Neutrino flares of radio blazars observed from TeV to PeV}",
      journal = {\mnras},
         year = 2024,
        month = jan,
       volume = {527},
       number = {1},
        pages = {L26-L31},
          doi = {10.1093/mnrasl/slad136},
archivePrefix = {arXiv},
       eprint = {2306.16797},
 primaryClass = {astro-ph.HE},
       adsurl = {https://ui.adsabs.harvard.edu/abs/2024MNRAS.527L..26S}
}

@ARTICLE{2023MNRAS.526..942A,
       author = {{Allakhverdyan}, V.~A. and {Avrorin}, A.~D. and {Avrorin}, A.~V. and {Aynutdinov}, V.~M. and {Barda{\v{c}}ov{\'a}}, Z. and {Belolaptikov}, I.~A. and {Bondarev}, E.~A. and {Borina}, I.~V. and {Budnev}, N.~M. and {Chepurnov}, A.~S. and {Dik}, V.~Y. and {Domogatsky}, G.~V. and {Doroshenko}, A.~A. and {Dvornick{\'y}}, R. and {Dyachok}, A.~N. and {Dzhilkibaev}, Zh-A.~M. and {Eckerov{\'a}}, E. and {Elzhov}, T.~V. and {Fajt}, L. and {Gafarov}, A.~R. and {Golubkov}, K.~V. and {Gorshkov}, N.~S. and {Gress}, T.~I. and {Kebkal}, K.~G. and {Kharuk}, I. and {Khramov}, E.~V. and {Kolbin}, M.~M. and {Konischev}, K.~V. and {Korobchenko}, A.~V. and {Koshechkin}, A.~P. and {Kozhin}, V.~A. and {Kruglov}, M.~V. and {Kulepov}, V.~F. and {Lemeshev}, Y.~E. and {Milenin}, M.~B. and {Mirgazov}, R.~R. and {Naumov}, D.~V. and {Nikolaev}, A.~S. and {Petukhov}, D.~P. and {Pliskovsky}, E.~N. and {Rozanov}, M.~I. and {Ryabov}, E.~V. and {Safronov}, G.~B. and {Seitova}, D. and {Shaybonov}, B.~A. and {Shelepov}, M.~D. and {Shilkin}, S.~D. and {Shirokov}, E.~V. and {{\v{S}}imkovic}, F. and {Sirenko}, A.~E. and {Skurikhin}, A.~V. and {Solovjev}, A.~G. and {Sorokovikov}, M.~N. and {{\v{S}}tekl}, I. and {Stromakov}, A.~P. and {Suvorova}, O.~V. and {Tabolenko}, V.~A. and {Ulzutuev}, B.~B. and {Yablokova}, Y.~V. and {Zaborov}, D.~N. and {Zavyalov}, S.~I. and {Zvezdov}, D.~Y. and {Kosogorov}, N.~A. and {Kovalev}, Y.~Y. and {Lipunova}, G.~V. and {Plavin}, A.~V. and {Semikoz}, D.~V. and {Troitsky}, S.~V. and {Baikal-GVD Collaboration}},
        title = "{Search for directional associations between baikal gigaton volume detector neutrino-induced cascades and high-energy astrophysical sources}",
      journal = {\mnras},
         year = 2023,
        month = nov,
       volume = {526},
       number = {1},
        pages = {942-951},
          doi = {10.1093/mnras/stad2641},
archivePrefix = {arXiv},
       eprint = {2307.07327},
 primaryClass = {astro-ph.HE},
       adsurl = {https://ui.adsabs.harvard.edu/abs/2023MNRAS.526..942A}
}

@ARTICLE{2023ApJ...955L..32B,
       author = {{Bellenghi}, Chiara and {Padovani}, Paolo and {Resconi}, Elisa and {Giommi}, Paolo},
        title = "{Correlating High-energy IceCube Neutrinos with 5BZCAT Blazars and RFC Sources}",
      journal = {\apjl},
         year = 2023,
        month = oct,
       volume = {955},
       number = {2},
          eid = {L32},
        pages = {L32},
          doi = {10.3847/2041-8213/acf711},
archivePrefix = {arXiv},
       eprint = {2309.03115},
 primaryClass = {astro-ph.HE},
       adsurl = {https://ui.adsabs.harvard.edu/abs/2023ApJ...955L..32B}
}

@INPROCEEDINGS{2023AAS...24111106D,
       author = {{Desai}, Abhishek and {Vandenbroucke}, Justin and {Pizzuto}, Alex and {Balagopal}, Aswathi and {Thwaites}, Jessie},
        title = "{Studying the AGN radio-neutrino correlation using IceCube and MOJAVE+RFC catalogs}",
    booktitle = {American Astronomical Society Meeting Abstracts},
         year = 2023,
       series = {American Astronomical Society Meeting Abstracts},
       volume = {55},
        month = jan,
          eid = {111.06},
        pages = {111.06},
       adsurl = {https://ui.adsabs.harvard.edu/abs/2023AAS...24111106D}
}

@ARTICLE{2023MNRAS.519.1396S,
       author = {{Sahakyan}, N. and {Giommi}, P. and {Padovani}, P. and {Petropoulou}, M. and {B{\'e}gu{\'e}}, D. and {Boccardi}, B. and {Gasparyan}, S.},
        title = "{A multimessenger study of the blazar PKS 0735+178: a new major neutrino source candidate}",
      journal = {\mnras},
         year = 2023,
        month = feb,
       volume = {519},
       number = {1},
        pages = {1396-1408},
          doi = {10.1093/mnras/stac3607},
archivePrefix = {arXiv},
       eprint = {2204.05060},
 primaryClass = {astro-ph.HE},
       adsurl = {https://ui.adsabs.harvard.edu/abs/2023MNRAS.519.1396S}
}

@ARTICLE{2023arXiv230713024R,
       author = {{Rodrigues}, Xavier and {Paliya}, Vaidehi S. and {Garrappa}, Simone and {Omeliukh}, Anastasiia and {Franckowiak}, Anna and {Winter}, Walter},
        title = "{Leptohadronic Multimessenger Modeling of 324 Gamma-Ray Blazars}",
      journal = {arXiv e-prints},
         year = 2023,
        month = jul,
          eid = {arXiv:2307.13024},
        pages = {arXiv:2307.13024},
          doi = {10.48550/arXiv.2307.13024},
archivePrefix = {arXiv},
       eprint = {2307.13024},
 primaryClass = {astro-ph.HE},
       adsurl = {https://ui.adsabs.harvard.edu/abs/2023arXiv230713024R}
}

@ARTICLE{2012ApJ...746...92C,
       author = {{Chandra}, Sunil and {Baliyan}, Kiran S. and {Ganesh}, Shashikiran and {Joshi}, Umesh C.},
        title = "{Optical Polarimetry of the Blazar CGRaBS J0211+1051 from Mount Abu Infrared Observatory}",
      journal = {\apj},
         year = 2012,
        month = feb,
       volume = {746},
       number = {1},
          eid = {92},
        pages = {92},
          doi = {10.1088/0004-637X/746/1/92},
archivePrefix = {arXiv},
       eprint = {1105.0572},
 primaryClass = {astro-ph.HE},
       adsurl = {https://ui.adsabs.harvard.edu/abs/2012ApJ...746...92C}
}

@ARTICLE{2010ApJ...712...14M,
       author = {{Meisner}, Aaron M. and {Romani}, Roger W.},
        title = "{Imaging Redshift Estimates for BL Lacertae Objects}",
      journal = {\apj},
         year = 2010,
        month = mar,
       volume = {712},
       number = {1},
        pages = {14-25},
          doi = {10.1088/0004-637X/712/1/14},
archivePrefix = {arXiv},
       eprint = {1002.1343},
 primaryClass = {astro-ph.CO},
       adsurl = {https://ui.adsabs.harvard.edu/abs/2010ApJ...712...14M}
}

@ARTICLE{2008ApJS..175...97H,
       author = {{Healey}, Stephen E. and {Romani}, Roger W. and {Cotter}, Garret and {Michelson}, Peter F. and {Schlafly}, Edward F. and {Readhead}, Anthony C.~S. and {Giommi}, Paolo and {Chaty}, Sylvain and {Grenier}, Isabelle A. and {Weintraub}, Lawrence C.},
        title = "{CGRaBS: An All-Sky Survey of Gamma-Ray Blazar Candidates}",
      journal = {\apjs},
         year = 2008,
        month = mar,
       volume = {175},
       number = {1},
        pages = {97-104},
          doi = {10.1086/523302},
archivePrefix = {arXiv},
       eprint = {0709.1735},
 primaryClass = {astro-ph},
       adsurl = {https://ui.adsabs.harvard.edu/abs/2008ApJS..175...97H}
}

@ARTICLE{2014ApJ...791...85C,
       author = {{Chandra}, Sunil and {Baliyan}, Kiran S. and {Ganesh}, S. and {Foschini}, L.},
        title = "{Understanding the Nature of the Blazar CGRaBS J0211+1051}",
      journal = {\apj},
         year = 2014,
        month = aug,
       volume = {791},
       number = {2},
          eid = {85},
        pages = {85},
          doi = {10.1088/0004-637X/791/2/85},
archivePrefix = {arXiv},
       eprint = {1406.7433},
 primaryClass = {astro-ph.HE},
       adsurl = {https://ui.adsabs.harvard.edu/abs/2014ApJ...791...85C}
}

@ARTICLE{2017MNRAS.464.4875B,
       author = {{Baring}, Matthew G. and {B{\"o}ttcher}, Markus and {Summerlin}, Errol J.},
        title = "{Probing acceleration and turbulence at relativistic shocks in blazar jets}",
      journal = {\mnras},
         year = 2017,
        month = feb,
       volume = {464},
       number = {4},
        pages = {4875-4894},
          doi = {10.1093/mnras/stw2344},
archivePrefix = {arXiv},
       eprint = {1609.03899},
 primaryClass = {astro-ph.HE},
       adsurl = {https://ui.adsabs.harvard.edu/abs/2017MNRAS.464.4875B}
}

\appendix

\section{{\large \onehale} code description} \label{sec:codes}

The \onehale\ code \citep{Zacharias21,zacharias+22} is a time-dependent, one-zone hadro-leptonic model calculating the particle distributions and photon spectra in a spherical region with radius $R$ permeated by a tangled magnetic field $B$. It moves with bulk Lorentz factor $\Gamma$, and as above, we assume here $\delta=\Gamma$. The code contains various options for external fields, such as the accretion disk, the broad-line region, the dusty torus, and the cosmic microwave background. However, in the application here, we only consider the broad-line region, the dusty torus, and the accretion disk as potential contributors to the IR-optical photon spectrum, but they do not partake strongly in the particle-photon interactions.

The particle distribution $n_i(\chi)$ of species $i$ (protons, charged pions, muons, and electrons, including positrons) with mass $m_i$ is given as a function of normalized momentum $\chi = p_i/m_i c = \gamma\beta$ with $\beta = \sqrt{1-\gamma^{-2}}$ and particle Lorentz factor \g. The distributions are derived from the Fokker-Planck equation,
\begin{align}
     \frac{\pd{n_i(\chi,t)}}{\pd{t}} &= \frac{\pd{}}{\pd{\chi}} \left[ \frac{\chi^2}{(a+2)t_{\rm acc}} \frac{\pd{n_i(\chi,t)}}{\pd{\chi}} \right] 
     - \frac{\pd{}}{\pd{\chi}} \left( \dot{\chi}_i n_i(\chi,t) \right)
	 - \frac{n_i(\chi,t)}{t_{\rm esc}} - \frac{n_i(\chi, t)}{\gamma t^{\ast}_{i,{\rm decay}}}
	 + Q_i(\chi,t)
	 \label{eq:z21_fpgen}.
\end{align}
The first term on the right-hand side describes Fermi-II acceleration through momentum diffusion employing hard-sphere scattering. The parameter $a$ is the ratio of the shock to Alv\`{e}n speed, while $t_{\rm acc}$ is the energy-independent acceleration time scale. The second term contains continuous energy gains and losses. The gain is Fermi-I acceleration described by $\dot{\chi}_{\rm FI} = \chi/t_{\rm acc}$, while the loss term contains the radiative and adiabatic processes of each particle species. All charged particles undergo synchrotron and adiabatic cooling, while protons additionally lose energy through Bethe-Heitler pair and pion production. We note that in the code version employed here, the conversion of protons to neutrons is not treated explicitly but is considered a continuous energy loss process instead. Electrons additionally undergo inverse-Compton losses, scattering all available internal and external photon fields.

The third term in Eq.~(\ref{eq:z21_fpgen}) marks the escape of particles described by t$_{esc}$=$\eta_{esc}R/c$, a multiple, $\eta_{esc}$, of the light travel time. The fourth term describes the decay of unstable particles, which decay in proper time $t^{\ast}_{i,{\rm decay}}$. The final term contains the injection of particles which is already described above. 

In the construction of this model, the Fermi-I and Fermi-II processes are merely used as re-acceleration mechanisms of the already relativistic particles in the shocked region with $t_{\rm acc} = \eta_{\rm acc}t_{\rm esc}$. The primary acceleration of particles might have taken place in a narrow sub-region around a sub-luminal mildly relativistic oblique shock, characterized as an acceleration zone \citep{2017MNRAS.464.4875B,2019ApJ...887..133B,Chandra_2021}. Here, the primary acceleration is approximated by an injection term Q$_{i;e^-/p^+}$($\chi,t$). Throughout this work, a simple power-law injection is used; the normalization of which is parameterized by
\begin{align}
    Q_{0,i}(t) = \frac{L_{inj,i}(t)}{Vm_ic^2}  
    \begin{cases}
        \frac{2-s_i(t)}{\gamma_{max,i}^{2-s_i(t)} - \gamma_{min,i}^{2-s_i(t)}}, & \text{if}\ s_i\ \neq 2 \\
        \Bigl( \ln\left(\frac{\gamma_{max,i}}{\gamma_{min,i}}\right) \Bigl), & \text{if}\ s_i\ = 2
    \end{cases}
    \label{eq:inj}
\end{align}
where s$_i$ refers to the spectral index of the particles (e$^-$/p$^+$) and V is the volume of the spherical emission region. The injection of secondary particles such as pions and muons are calculated directly from the photo-hadron interactions \citep{Hummer_2010} and their decay. 
Secondary electrons are injected from muon decay, Bethe-Heitler pair production, and \g-\g\ pair production. Neutral pions decay quickly into \g\-rays, which is why the resulting radiation is computed directly from their injection spectrum.

\begin{align}
    \frac{\pd{n_{ph}(\nu, t)}}{\pd{t}} &= \frac{4\pi}{h\nu}j_\nu(t) - n_{ph}(\nu,t)\left(\frac{1}{t_{esc,ph}} + \frac{1}{t_{abs}}\right)
    \label{eq:contequ}
\end{align}
As Eq.~(\ref{eq:z21_fpgen}) relies on the time-dependent photon spectrum present within the emission region, in each time step, the Fokker-Planck equation is solved for all charged particle species along with the radiation transport equation, Eq.~(\ref{eq:contequ}). This setup ensures that the updated photon distribution is used for the next time-step for all particle-photon and photon-photon interactions. 
The first term of the equation \ref{eq:contequ} is the photon production term parameterized by the emissivity and the second term reflects the combined losses of photons due to escape and absorption. The emissivity includes contributions from the synchrotron and inverse-Compton processes. 

While the code is fully time-dependent, it can be used to calculate steady-state solutions as well. This is achieved if the total particle densities of protons $n_p$ and electrons $n_e$ each vary less than $10^{-4}$ relative to the two previous time steps. The detailed equations of the whole code can be found in \cite{zacharias+22}.

\end{document}